# Towards the superlubricity of polymer–steel interfaces with ionic liquids and carbon nanotubes

Ł. Wojciechowski* [a)], K.J. Kubiak [b)], S. Boncel [c) d)], A. Marek [c) d) e)], B. Gapiński [f)], T. Runka [g)], R. Jędrysiak [c) d)],
S. Ruczka [c) d)], P. Błaszkiewicz [h)], T.G. Mathia [i)]

[a)] Institute of Machines and Motor Vehicles (IMRiPS), Poznan University of Technology, Poland
[b)] School of Mechanical Engineering, University of Leeds, United Kingdom
[c)] Centre for Organic and Nanohybrid Electronics, Silesian University of Technology (CONE), Poland
[d)] Department of Organic Chemistry, Bioorganic Chemistry and Biotechnology, Silesian University of Technology, Poland
[e)] Department of Chemical Organic Technology and Petrochemistry, Silesian University of Technology, Poland
[f)] Division of Metrology & Measurement Systems, Poznan University of Technology, Poland
[g)] Institute of Materials Research and Quantum Engineering, Poznan University of Technology, Poland
[h)] Institute of Physics, Poznan University of Technology, Poland
[i)] Laboratoire de Tribologie et Dynamique des Systèmes – CNRS, École Centrale de Lyon, France

**Abstract**

Frictional losses are responsible for significant energy waste in many practical applications, and superlubricity with a coefficient of friction lower than 0.01 is the goal of tribologists. In this paper, metal-on-polymer contact was analysed and close to superlubricity conditions for this material configuration were explored. A new lubricant has been proposed hinge on the phosphorus-based ionic liquid and carbon nanotubes as thickeners. Additionally, carbon nanotube mesh was doped with copper nanoparticles that allowed for the close to superlubricity state in a mild steel/polymer contact configuration under low normal load conditions. The adsorption of phosphorus onto metallic and polymer surfaces has been reported in EDS analysis. The formulation of the new lubricant allowed for stable dispersion with a carbon nanotube content as low as 0.1% wt. The carbon nanotubes and Cu nanoparticles have been analysed using TEM and SEM imaging. A tribological test in a block-on-ring system has been carried out. The wear of material, topography, and surface free energy have been analysed along with SEM/EDS images to explore the underlying mechanisms of friction and wear.

| Nomenclature: | |
|---|---|
| AFC | Abbott-Firestone curve |
| CNTs | Carbon nanotubes |
| COF | Coefficient of friction |
| EDS | Energy Dispersive Spectroscopy |
| IL | Ionic liquid |
| MWCNTs | Multi-walled carbon nanotubes |
| NCs | Nanocarbons |
| SEM | Scanning Electron Microscopy |
| SFE | Surface free energy |
| SWCNTs | Single-walled carbon nanotubes |
| TEM | Transmission Electron Microscopy |
| $V_{mc}$ | Core material volume [ml/m$^2$] |
| $V_{mp}$ | Peak material volume [ml/m$^2$] |
| $V_{vc}$ | Core void volume [ml/m$^2$] |
| $V_{vv}$ | Valley void volume [ml/m$^2$] |

## 1. Introduction

One of the greatest challenges for humanity in the 21st century is to reduce the consumption of fossil fuels and related $CO_2$ emissions into the atmosphere. Among the various technological solutions leading to this goal is the reduction of energy losses related to friction and wear. The development of innovative formulations of lubricants seems to be one of the trends that lead to such an effect, allowing us to achieve a state in which the value of the friction coefficient drops clearly below 0.01. This state is called superlubricity and is defined as a lubrication regime in which the adhesive interactions between the rubbing surfaces are so low that they are considered negligible. Many mechanisms are responsible for achieving superlubricity. This applies to both the so-called liquid lubricity and solid superlubricity. Detailed reviews on this issue were presented by Han et al. [1] and Ayyagari et al. [2]. Initially, the super-lubricating effect was obtained mainly on the nano/micro scale, mainly through the use of electron microscopes [e.g. 3-5]. However, recently there have been more studies in which superlubricity was obtained on a macro scale, during tests on classic tribometers. Chen et al. [6] showed that polymer-based ionic liquids (ILs) applied to lubricate a steel–steel friction pair can achieve a superlubricity effect even at high pressures (above 1 GPa) and without a running-in phase. Xu et al. [7] obtained the effect of superlubricity by using a synthetic 1,3-diketone enriched with silica nanoparticles to lubricate the friction pair of the bearing steel. However, the authors point out that obtaining this effect depends on specific test conditions (relationship of speed and load) and the potential antagonism of such a lubricant to traditional anti-wear additives (e.g. ZDDP, $MoS_2$). In turn, Ren et al. [8] suggest that superlubricity can also be obtained for traditional tribological additives, provided that the lubricating film between the rubbing surfaces is properly composed. For this purpose, they designed superlattice lubricating films consisting of alternating molybdenum disulphide and tungsten carbide layers. This arrangement allowed for a very low COF (0.006) under light vacuum conditions. Wang et al. [9] studied the effect of ultra-low friction on ceramic interfaces lubricated with imidazole ILs. For such test assumptions, superlubricity conditions were obtained for long-term tests under high loads. The potential of ionic liquids in the creation of superlubricants was also noticed in the recent papers: [10] (PTFE lubricated by protic ILs), [11] (lubrication with a water solution of glycerol with the addition of ionic liquids), and [12] (synergy of ILs and graphene oxide nanosheets in a ceramic interface). Another way to achieve the superlubrication effect is to add nano-additives to lubricants. In this concept, nanocarbons – which in various forms (depending on dimensionality, from 0D to 2D) are being investigated as potential tribological additives to oils and greases [13] – seem to dominate. The nanocarbons (NCs) most often analysed in the context of anti-wear applications include carbon nanotubes (CNTs), discovered by Iijima [14]. There are two types of CNTs: multi-walled (MWCNTs) and single-walled (SWCNTs). MWCNTs can be defined as several layers of concentrically arranged plates of graphitised carbon. SWCNTs can be described more as a single sheet of graphene rolled into a tube [15]. Regardless of the type, CNTs are characterised by a diameter of several to tens of nm and a length of several to several tens of μm – on this basis, they are classified as 1D NCs. A characteristic feature of CNTs is very strong van der Waals interactions between them. This causes their tendency to form agglomerates, which prevents their solubility in most solvents and makes it difficult to disperse them in liquids. To avoid these problems, CNTs are subjected to functionalisation, which is carried out by covalent attachment of desired functional groups with the existing structure of CNTs, or by non-covalent adsorption or 'wrapping' of (macro) molecules on the tube surfaces [15]. The functionalisation of CNTs allows them to have the desired, often application-dependent, additional properties. Such an effect can be obtained by decorating CNTs with other nanoparticles, which, first, will hinder their mutual agglomeration, and secondly, will introduce new, beneficial properties to their structure [16]. Various examples of application of this solution can be found in medical diagnostics [17-18], comprising antibacterial remedies [19], or the production of multi-functional composites [20-21]. These possibilities have also been noticed concerning the possibility of improving the tribological properties of CNTs. As for the idea, it is possible to enrich the structure of CNTs with metal nanoparticles whose anti-friction and anti-wear effects as an additive to lubricants have been confirmed. For example, such an effect was obtained for copper [22-23], silver [24-25] and nickel [26-27] nanoparticles. The expected effect of synergism between CNTs and metal nanoparticles used for their decoration was identified by Meng et al. for silver [28] and nickel [29]. What is extremely important here, in both studies, is that the oil enriched with both CNTs and metal nanoparticles provided lower COF and wear than oils with only CNTs or with only Ni or Ag nanoparticles as per weight ratio.

*Corresponding author: Lukasz Wojciechowski, Institute of Machines and Motor Vehicles, Poznan University of Technology,
ul. Piotrowo 3, 60-965 Poznan, Poland, e-mail: Lukasz.Wojciechowski@put.poznan.pl

NCs' high reactivity and the resulting potential for synergism with other surface-active compounds are also used to produce tribo-additives for oils and greases. For example, Gan et al. [30] investigated the use of hydroxyl-terminated ILs as a functionalising agent for graphene oxide as a water-based oil additive. This configuration had a positive effect on the dispersion of graphene and the lubricity of the final lubricant (although not at the level of superlubricity). Similar conclusions result from the work of Bo et al. [31] who tested the anti-wear properties of imidazolium-based IL–MWCNTs formulations as additives to model lubricants. The authors found a clearly beneficial effect in reduced friction and wear, as well as good dispersibility immediately after sonication. However, keeping the mixture in a steady state for some time (5 h) indicated its sedimentation potential. This problem, i.e. long-term dispersion stability, is one of the factors that hinders the practical application of lubricants containing CNTs. The solution may be the use of surfactants that will help suspend nanoparticles in the liquid phase of the lubricant and thus maintain their homogeneous dispersion. Various surfactants are known to be used for this purpose, e.g. SDS (sodium dodecyl sulphate) [32-33] and octyl phenol ethoxylate [33].

When analysing the use of NCs in lubricant formulations, the multi-threaded nature of the issue draws attention. The prevailing opinion is that tribo-active carbon nanoparticles have a beneficial effect on reducing friction and wear (particularly in metallic pairs). However, they require appropriate functionalisation, which may have a very different nature depending on the intended use. This results in the lack of a clear view of the mechanisms of interaction of NCs with metal or polymer surfaces and a universal protocol on how to use them as an additive to oils or greases.

Considering the above problems, in this paper, we would like to present our novel view on applying these intriguing tribo-active additives for forming lubricants that allow low COF values to be achieved even for relatively high loads. For this purpose, we propose the creation of a hybrid grease using the excellent anti-friction properties of both ionic liquid and CNTs, applied as a liquid phase and thickener, respectively. They were used to lubricate polymer–steel friction pairs to compare their impact on reducing friction and wear with that of a commercially available reference grease. An analysis of wear traces on polymer and steel surfaces was also performed to identify wear mechanisms and propose potential processes for creating protective films with hybrid lubricants.

**2. Materials and methodology**

2.1. Preparation of hybrid lubricants

A new hybrid type of lubricant has been prepared in this study, composed of ionic liquid and thickener phases. Industrial grade MWCNTs: NC7000™, ~90 wt% carbon, ~9 wt% $Al_2O_3$ support and ~1 wt% iron-based catalyst (Nanocyl Co. Ltd, Sambreville, Belgium) were selected as a base material for the creation of thickener. According to the manufacturer's specification, these MWCNTs are 1.5 µm long and 9.5 nm in diameter. The presence of $Al_2O_3$ is a catalyst support for the catalytic vapour deposition (CVD) growth of MWCNTs. Unfortunately, $Al_2O_3$ is characterised by high hardness, and its presence, even on the nanoscale, may have a very negative impact on wear mechanisms in the contact zone [34]. Therefore, before the process of combining with the liquid phase, the MWCNTs were purified to remove $Al_2O_3$ [45]. SEM images of MWCNTs before and after purification are shown in Figure 1. Importantly, as one might expect from the conditions applied (e.g. possible oxidation via OH radicals generated in situ from hydroxyl anions), the purification process had an impact on the surface chemistry and morphology of the nanotubes. And so, the purified MWCNTs contained 0.45±0.10 mmol COOH $g^{-1}$ and 0.58±0.21 mmol OH $g^{-1}$ according to Boehm titration [67]. Such purification, inherently interrelated with a partial functionalisation, caused formation of a hydrogen-bonded 3D network which was observed macroscopically as increased rigidity of the nanotube cake formed upon being filtered off. The overall increase in compactness led to a decrease of BET specific surface area by c. 70 $m^2/g$ (from 300 to 213 $m^2/g$) that, nevertheless, should be taken cautiously [68].

One of the options for preparing lubricants was to use a nanotube thickener decorated with copper nanoparticles. Cu-CNTs were synthesised via a modified polyol reduction method [35]. First, purified MWCNTs (0.25 g) were dispersed by ultrasonication in ethylene glycol (100 mL, used as the solvent as well as a reducing and protecting/stabilising agent). Then, sodium hydroxide (1.20 g) and the metal precursor (5 mmol), i.e. copper(II) nitrate trihydrate, were added to the mixture. After the complete dissolution of the precursor, chloroplatinic acid solution (250 µL, 8 wt%) was added to the mixture as the heterogeneous nucleation seeds. Subsequently, the mixture was heated using a hot-plate magnetic stirrer, and the reduction was carried out at 100 °C for 4 h under stirring. After cooling to room temperature, the Cu-CNT solid was filtered off and washed thoroughly with water and ethanol. The resulting catalyst was dried at 60 °C in a vacuum oven for 24 h.

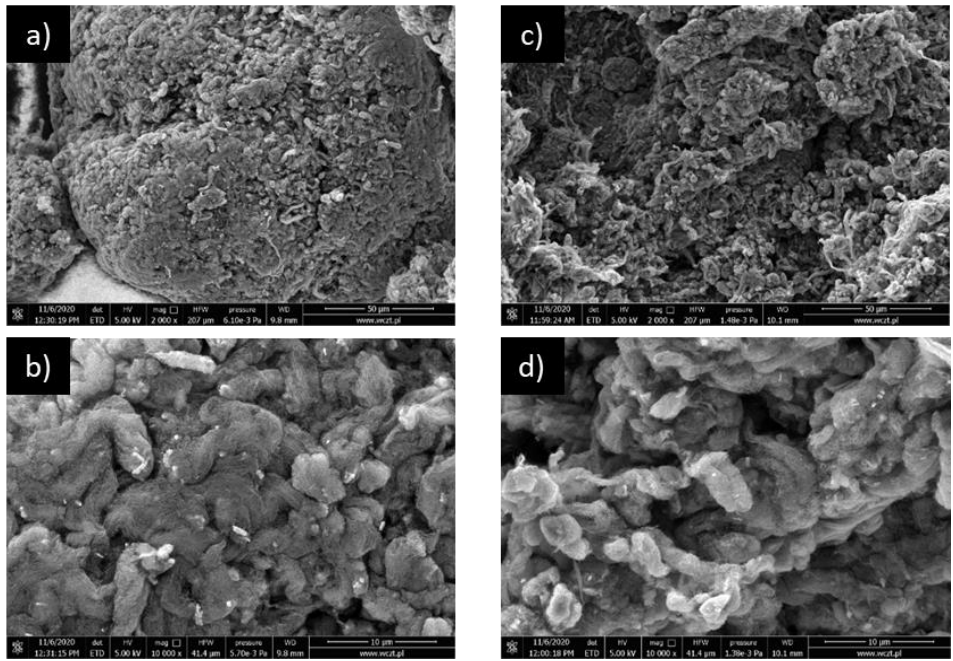

Figure 1. SEM pictures of CNTs before (a – mag. 2000x, b – mag. 10000x) and after (c – mag. 2000x, d – mag. 10000x) purification.

Figure 2 shows TEM pictures of separate multi-grain copper nanoparticles embedded into spaghetti-like CNT bundles. It should be noted that passivation-derived hydroxyl groups are an important cofactor of the metallic Cu nanoparticles owing to a high surface area and the latter's increased reactivity toward the environmental oxygen and water. This structure could allow the formation of hydrogen bonds between dangling surface hydroxyl groups of Cu nanoparticles and COOH/OH groups from the purified MWCNTs [65,66]. In general, two types of nanotube network connections can be identified here. Flexible, highly entangled structures (green) dominate, consisting only of CNTs and constituting the basis of the three-dimensional thickener network. Additionally, crystalline direct joints of nanoparticles (yellow) are visible, which are not a consequence of CNT entanglement. The original presence of Cu agglomerates would enable homogenisation of the Cu nanoparticles among all the components of the tribopair system at the nanoscale, with the higher performance discussed in detail later.

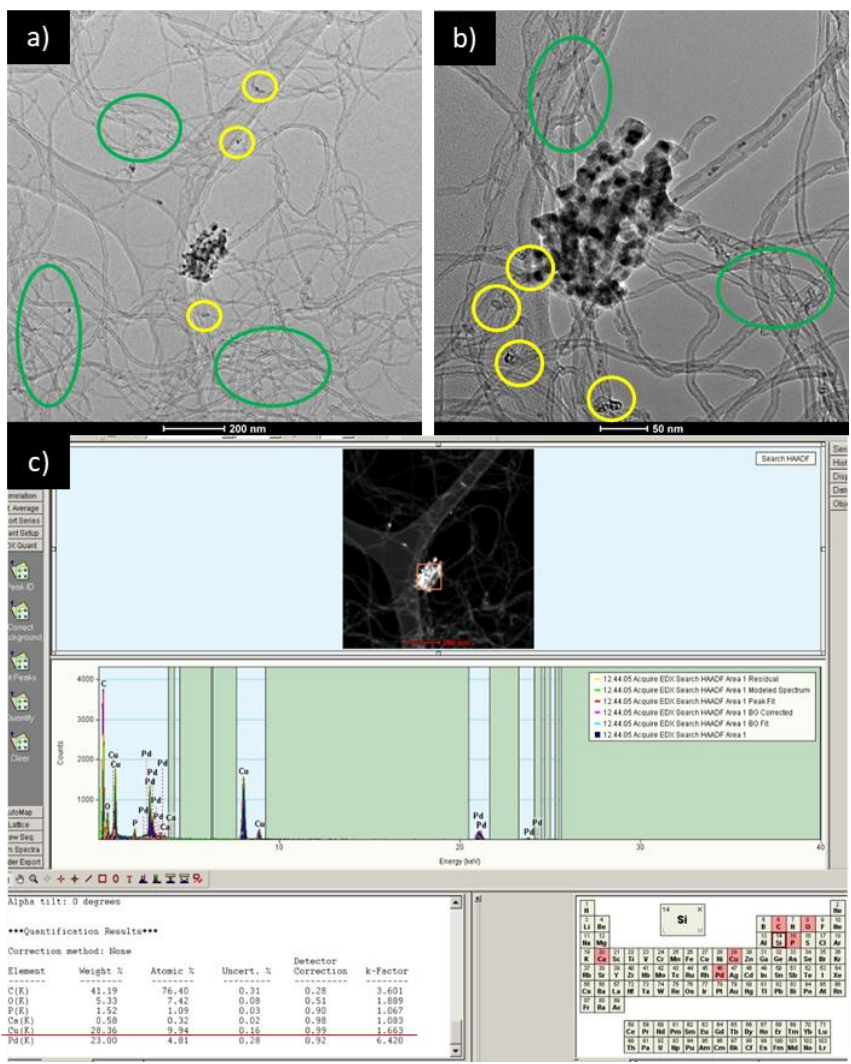

Figure 2. TEM pictures (a, b) of separate multi-grain copper nanoparticles embedded into spaghetti-like CNT bundles with examples of two types of network connections: highly entangled CNTs' structures (green), crystalline, direct joints of nanoparticles (yellow); an exemplary local surface scan – EDX spectrum of Cu-CNT nanohybrid (c).

To investigate the possible impact of purification and decoration with copper on the structural order of CNTs, Raman analysis was performed. Raman spectra were recorded using a Renishaw inVia Raman microscope equipped with a thermoelectrically (TE)-cooled CCD detector, an Ar$^+$ ion laser working at a 514.5 nm wavelength. The spectra were recorded in the spectral range 100–3,500 cm$^{-1}$ with a spectral resolution better than 2 cm$^{-1}$. The power of the laser beam focused on the sample with an ×50 objective was kept below 1 mW. The positions of peaks were calibrated before collecting the data, using crystalline silicon. The spectral parameters of the bands were determined after baseline correction using the fitting procedure of the Wire 3.4 software. Figure 3 presents the Raman spectra of CNTs with $Al_2O_3$, CNTs after purification and CNTs after purification and decoration with Cu, recorded upon excitation with a 514.5 nm wavelength.

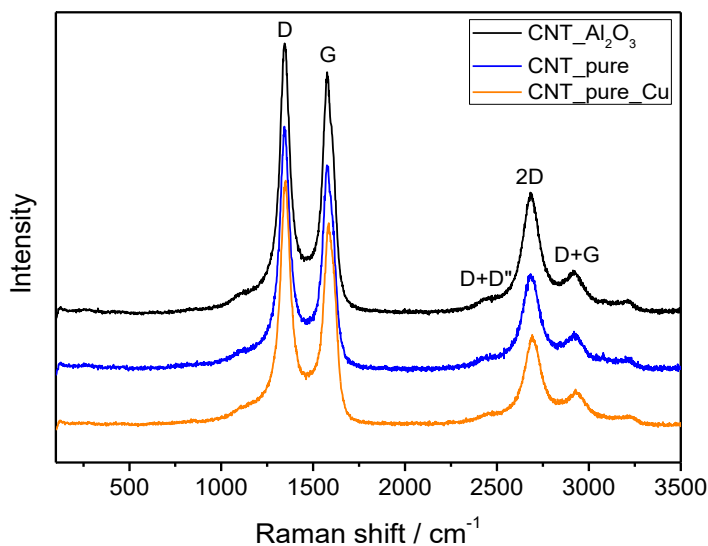

Figure 3. Raman spectra of CNTs with $Al_2O_3$, CNTs after purification and CNTs after purification and decoration with Cu, recorded in the spectral range 1,000-3,500 $cm^{-1}$ upon excitation with a 514.5 nm wavelength.

In carbon allotropes containing a mixture of $sp^2$ and $sp^3$ type carbon-carbon bonding, there are two important characteristic Raman bands at 1,580 $cm^{-1}$ and 1,350 $cm^{-1}$, which were assigned to the in-plane vibrations of the C – C bonds (G bands, stretching of all $sp^2$ bonds, both in rings and chains) and vibration mode originated from the distorted hexagonal lattice of graphitic $sp^2$ network near the crystal boundary (D bands), respectively. The insignificant peak at around 2,460 $cm^{-1}$ (D + D") is related to a combination of a D phonon and an acoustic longitudinal phonon D". At higher wavenumbers 2D (G') and D + G band at around 2,690 $cm^{-1}$ and 2,920 $cm^{-1}$ were observed, respectively [53]. It has been shown that the relative intensity ratio ($I_D/I_G$) of the D to the G band is an indicator of the perfection of the graphite layer surface. In general, the increase in the $I_D/I_G$ ratio indicates an increased number of defects of the carbon species [54]. Table 1 shows the comparison of $I_D/I_G$ ratio for CNTs with $Al_2O_3$, CNTs after purification and CNTs after purification and decoration with Cu. High $I_D/I_G$ values within the range 1.1–1.4 for investigated CNTs indicate numerous graphitisation defects, especially in the outermost MWCNT walls. The $I_D/I_G$ ratio can also be used to estimate a planar domain size $L_a$ in carbon allotropes [55]. Table 1 shows that $L_a$ decreases slightly from 15.0 nm to 12.2 nm, confirming that purification and functionalisation do not change the structure of CNTs significantly.

Table 1. The $I_D/I_G$ ratio and domain size $L_a$ for CNTs with $Al_2O_3$, CNTs after purification and CNTs after purification and decoration with Cu.

| CNTs type | $I_D/I_G$ | $L_a$ (nm) |
|---|---|---|
| CNTs with $Al_2O_3$ (CNT_$Al_2O_3$) | 1.119 | 15.0 |
| CNTs after purification (CNT_pure) | 1.255 | 13.4 |
| CNTs after purification and decoration with Cu (CNT_pure_Cu) | 1.380 | 12.2 |

Trihexyltetradecylphosphonium bis(2-ethylhexyl) phosphate (P666 DEHP (14), $C_{48}H_{102}O_4P_2$) was selected as a liquid phase. This ionic liquid is a well-known oil additive with proven anti-friction, anti-wear and anti-corrosion properties [36-41]. However, its application as a liquid phase for lubricants has not been explored so far.

To obtain a homogeneous dispersion of MWCNTs in IL, the homogenisation method was applied (Unidrive X1000 Homogeniser with a G20 cutting knife tip). Samples were subjected to shearing forces at 20,000 rpm in 3 cycles of 30 s (Figure 4). Four lubricant compositions with different CNT contents were prepared according to the previously described procedure (Figure 4). These were: IL+CNT1%, IL+CNT0.5%, IL+CNT0.1%, IL+CNT_Cu0.1%.

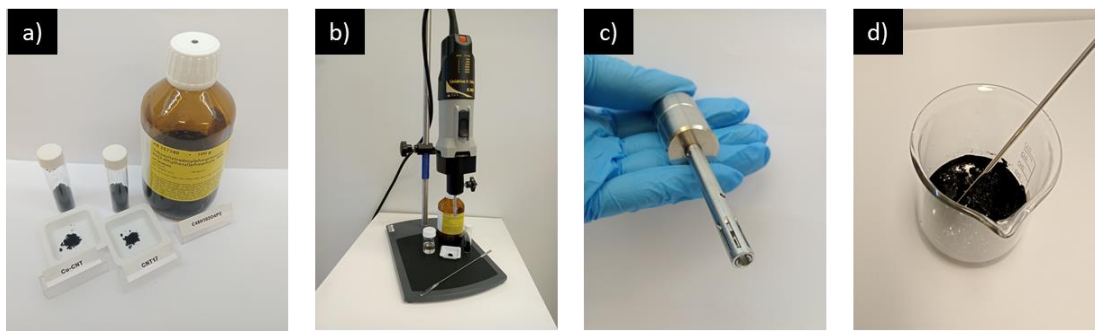

Figure 4. Ingredients and equipment for producing hybrid lubricants: selected ionic liquid and CNTs (a), high-shear mixer (b), knife for low-volume samples (c), ready-made lubricant (d).

The dynamic viscosity of the prepared lubricants was tested using a Brookfield viscometer under the following conditions: temperature 40°C, spindle speed 5 rpm (shear rate 1.7 1/s), and 10 measurements for each lubricant. The average results with standard deviations are presented in Table 2. As could be expected, the increase in the content of CNTs in the lubricant resulted in an increase in its viscosity. The exception to this rule is the lubricant with the addition of Cu-decorated CNTs, for which no change in viscosity was found compared to pure IL.

Table 2. Average dynamic viscosity of prepared lubricants (temp. 40°C, spindle speed 5 rpm, shear rate 1.7 1/s).

| Lubricant composition | Dynamic viscosity [mPa·s] |
| --- | --- |
| $C_{48}H_{102}O_4P_2$ | 1077 ± 7 |
| $C_{48}H_{102}O_4P_2$+Cu_CNT0.1% | 1078 ± 38 |
| $C_{48}H_{102}O_4P_2$+CNT0.1% | 1794 ± 22 |
| $C_{48}H_{102}O_4P_2$ + CNT0.5% | 3169 ± 52 |
| $C_{48}H_{102}O_4P_2$ + CNT1% | 5804 ± 83 |

## 2.2. Tribological tests and wear investigations

Tribological tests were carried out with the application of the 'block-on-ring' system (Figure 5), on a modernised Amsler A135 tribometer. Cylindrically ground (Sa ≈ 0.5 μm) rings with an external diameter of 45 mm and width of 12 mm were manufactured from AISI 4130 steel. Polymer blocks (15 x 10 x 6 mm) were made of commercially available materials for friction applications: UHMWPE (trade name: Tivar 1000), POM-C (Ertacetal C) and PA-6 (Ertalon 6SA). Their contact surfaces (15 x 10 mm) were ground to obtain the same roughness. The tests were performed under kinetic conditions: ring rotation speed of 200 rpm, load of 1,000 N, duration of 30 min. The load application diagram is shown in Figure 5. All configurations of friction pairs were lubricated with the same volume of lubricant – 0.1 mL. Frictional behaviour was analysed using the measured values for the friction torque and the COF calculated from it (and the applied load) for one-minute intervals. For statistical purposes, tests were repeated three times for each configuration of friction pair materials and lubricants. After each test, excess lubricant was removed from the working surfaces of the blocks and cylinders. However, these surfaces were not additionally chemically cleaned before wear analysis.

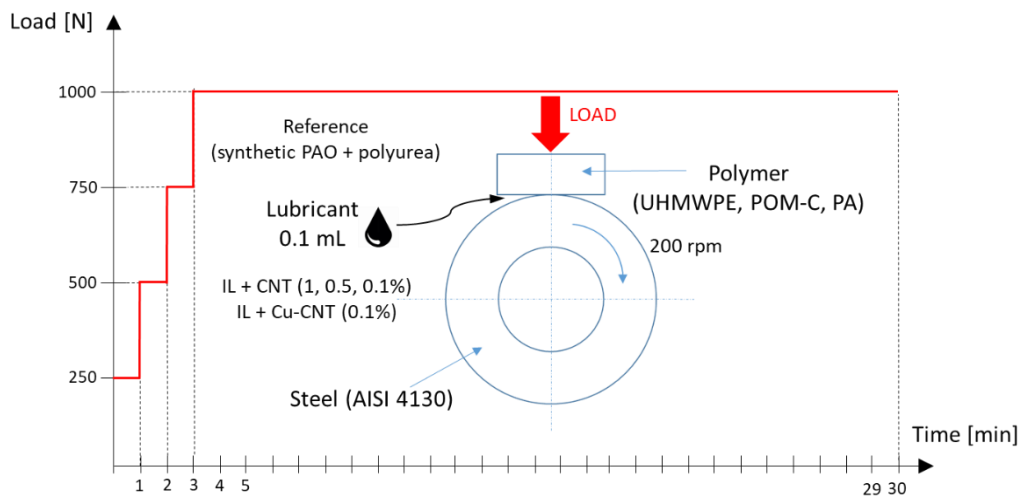

Figure 5. Materials configurations and kinetics (including a load application diagram) of tribological tests.

SEM and EDS investigations were carried out for each type of polymer on randomly selected working surfaces of new samples and on traces of wear on the blocks after the tribological tests. The same methodology was used for steel rings – the working surfaces of samples cooperating with each type of polymer used in the tests were analysed. Based on the observations from these investigations, the wear mechanisms of the polymer and steel samples were recognised and characterised.

The areal topographic analysis was carried out by applying the 3D contact profilometer to the new and worn surfaces of polymer samples. The surface topography analysis was performed on standardised (ISO25178) height and volumetric parameters.

The surface free energy (SFE) of polymers was calculated based on the results of contact angle measurements performed with model liquids (water, di-iodomethane) on new and worn surfaces. The Attension Biolin goniometer was used for the wettability measurement. The test was repeated three time for each surface. The Owens–Wendt–Rabel–Kaelble (OWRK) methodology was used to determine the total value of SFE and its dispersive and polar parts [42].

## 3. Results and discussion

The results and discussion section is divided into friction, wear, topography and SFE analysis.

### 3.1. Friction tests

Figures 6–8 show averaged COF trends with standard errors calculated for one-minute intervals, for all analysed polymers. The results of friction torque measurements for the same time intervals are summarised in Appendix A. In the first case considered (AISI4130-UHMWPE), it can be observed that two of the lubricants used offer significant reduction in the friction coefficient compared with the reference grease. These are IL with 1% CNTs and IL with 0.1% Cu-CNTs. In the initial phase (1 min of test), their COF values are only slightly lower (IL + CNT1%→0.013, IL + Cu-CNT0.1%→0.012) than the reference grease (0.014). However, if we consider the COF values in the final phase of the test (e.g. 30 min, load 1kN), we notice that the COF values for these lubricants are much lower than for the reference grease. For IL + CNT1%, the COF is equal to 0.029, for IL + Cu-CNT0.1%, and for the reference, it is as much as 0.041. The fact that these two lubricants have the fastest friction stabilisation is also a very promising sign. After reaching the time interval of approx. 720-840 s, the COF value is set at a similar level. For the reference grease, no such stabilisation was observed, but after approximately 800 s of the test, the COF oscillates until the end in the range of 0.035-0.041. The remaining lubricants (IL + CNT0.5% and IL + CNT0.1%) show similar, or poorer friction behaviour than the reference grease.

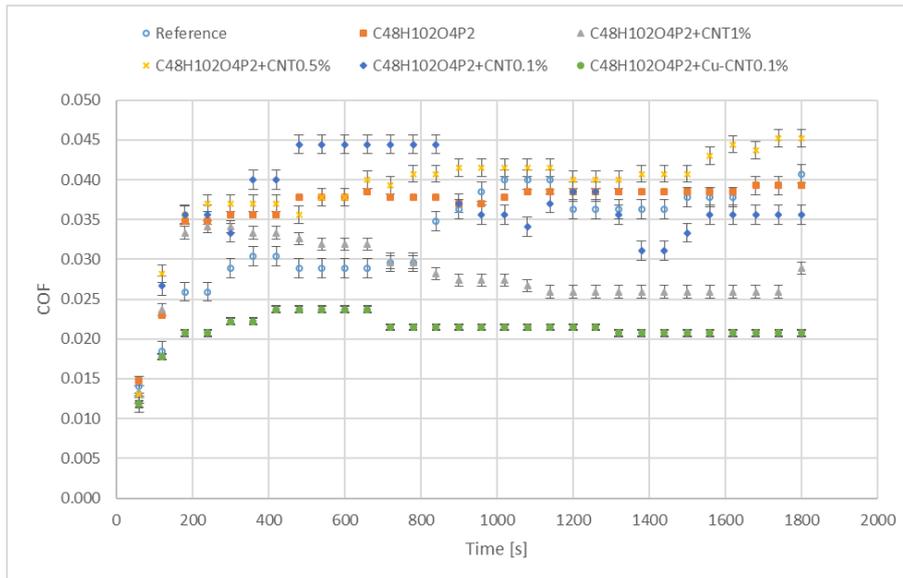

Figure 6. The average trends of COF for the AISI4130–UHMWPE friction pair.

In the second case, the AISI 4130–POM-C friction pair (Figure 7), we see that none of the hybrid lubricants (except IL + Cu-CNT0.1%) provides a lower COF than the reference grease. This trend occurs both in the initial phase and after the stabilisation of the friction. What is very interesting here is that almost all lubricants stabilise the operation of the friction pair relatively quickly. This takes approximately 400 s. It is also worth noting that for most lubricants, COF is characterised by higher values than the AISI4130 with UHMWPE configuration (Figure 6). The exception is the reference grease, for which COF is clearly lower (1800 s → 0.032) for POM-C than for the previously analysed UHMWPE (1800 s → 0.041). Therefore, it can be concluded that neither pure IL nor the addition of CNTs to IL had a sufficiently beneficial effect on reducing friction for this material configuration. The real 'game changer' turned out to be the addition of Cu to CNTs. For IL + Cu-CNT0.1%, the COF value is the lowest at each stage of the test.

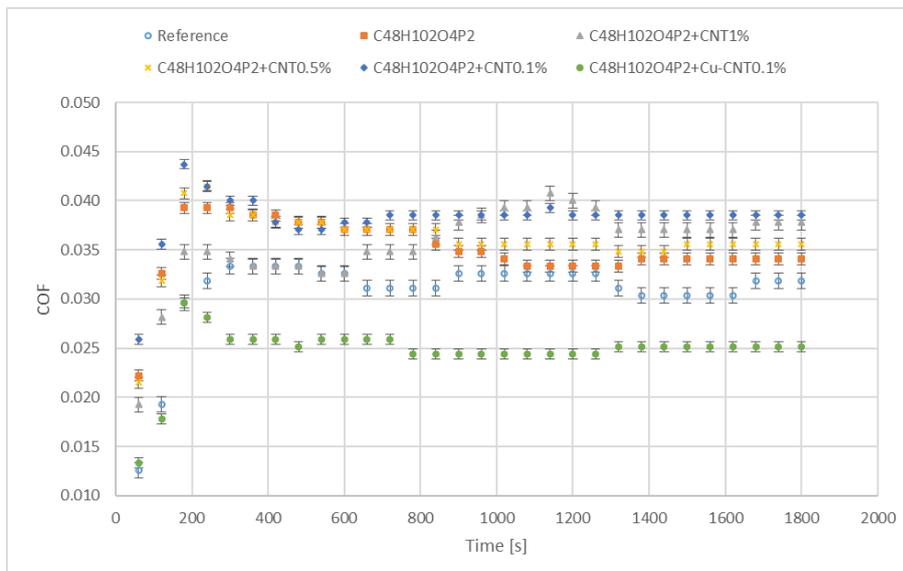

Figure 7. The average trends of COF for the AISI4130–POM-C friction pair.

The third case is related to friction tests for the AISI 4130–PA pair (Figure 8). One can observe here similar trends as those identified for the first material configuration (AISI 4130–UHMWPE → Figure 6). This means that the lowest COF values were achieved for lubrication with two lubricants, i.e.: IL + CNT1% (0.026 in 1800 s) and IL + Cu-CNT0.1%. (0.033 in 1800 s). The trend of the initial phase of the test is very interesting here. As can be seen in Figure 8, for the lowest of the applied loads, the COF value for the Cu-decorated CNTs lubricant achieved extremely low values. In the first 60 s of the test, Cu rich configuration had COF value of only ~0.009, while after 120 s of the test, its value increased to 0.013. Therefore, it is justified to conclude that conditions close to superlubricity can be short-term achieved for the steel-polyamide pair under low load when Cu nanoparticles are present in the contact. As with alternative polymers, both lubricants quickly stabilised the friction pair operation. COF values reached a similar level after approximately 700 s of the test duration. Also, the lower content of undecorated CNTs in IL ensured a reduction in friction in this case compared to the reference grease. It should be noted, however, that ILs with 0.5% and 0.1% CNTs content are characterised by very similar change trends and COF values. All of them are clearly lower than the reference grease. Therefore, for the lubrication of a friction pair in which a polyamide part is used, lubricants based on phosphate ILs and CNTs (including those decorated with Cu) have great potential for reducing friction. It is also possible to recognise the relationship between the CNTs' content and the friction reduction potential. The higher the amount of CNTs in the lubricant, the lower the COF values observed during the tests. Of course, this conclusion holds only within the range of CNTs content in IL used (from 0.1 to 1%).

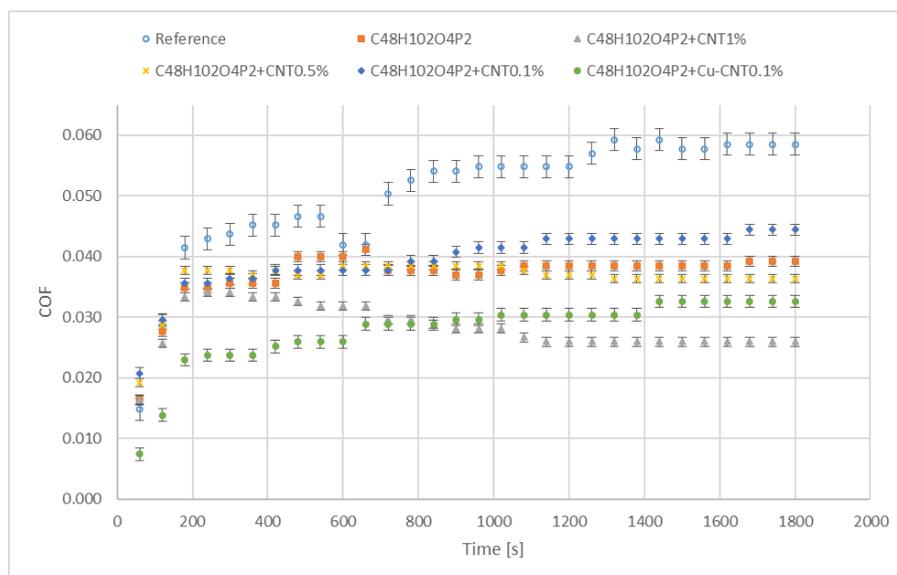

Figure 8. The average trends of COF for the AISI4130–PA friction pair.

3.2. Wear mechanisms

The interpretation of the wear mechanisms was based on SEM/EDS investigations and topographic measurements for selected test configurations. In this part of the experiment, the focus was mainly on samples after friction tests, in which the lubricant was IL + Cu-CNT0.1%. This is due to the innovative composition of this lubricant and its very beneficial effect on the friction behaviour of all tested material configurations. Figures 9 and 10 present SEM pictures and EDS analysis of wear signs on surfaces of UHMWPE blocks and corresponding steel rings lubricated by IL + Cu-CNT0.1%.

Observing wear traces on the UHMWPE blocks allows the recognition of small material losses, probably caused by micro-adhesion. This is a typical form of polymer wear in contact with metal materials, the source of which is the so-called 'lumpy transfer' [43]. During the plastic flow of polymers under friction conditions, favourable conditions are created for the formation of adhesive joints of this material with metal. As a result, small parts of the polymer are torn out from the surface and transferred onto the metal, on which a polymer film is formed. Generally, this is a very beneficial phenomenon that generates the formation of polymer-on-polymer interphase, in which the friction is usually milder than direct polymer-on-metal contact. However, the problem is that the polymer film on the metal surface is usually lumpy and has a random, heterogeneous structure. This means that only part of the contact surface has a polymer interphase, and some of it remains in metal-on-polymer contact.

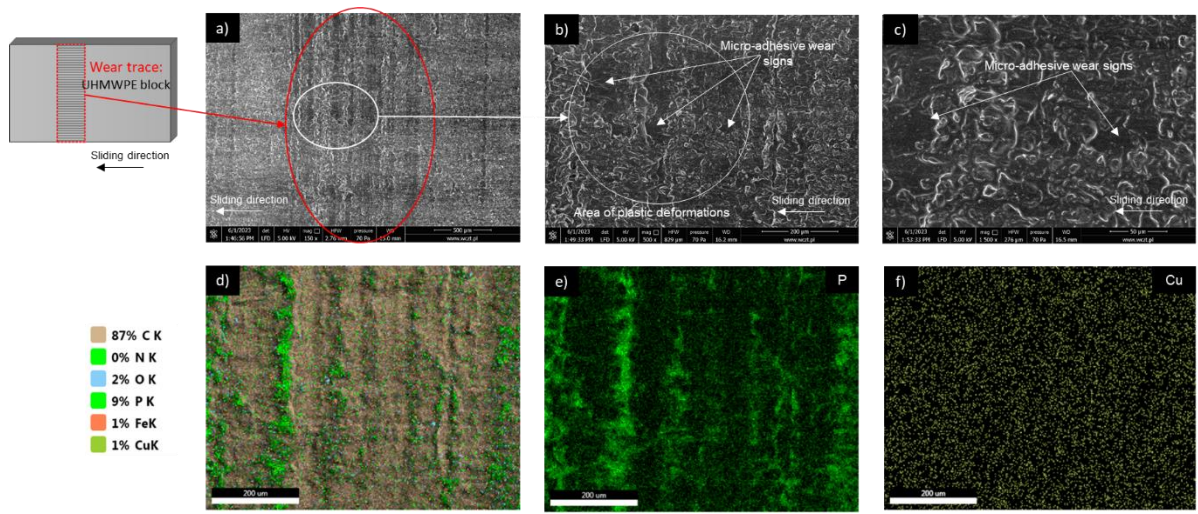

Figure 9. Investigations of wear signs on UHMWPE blocks lubricated by IL + Cu-CNT0.1%: SEM pictures with different magnifications: 150x (a), 500x (b), 1500x (c), EDS analysis – general view (d), P content (e), Cu Content (f).

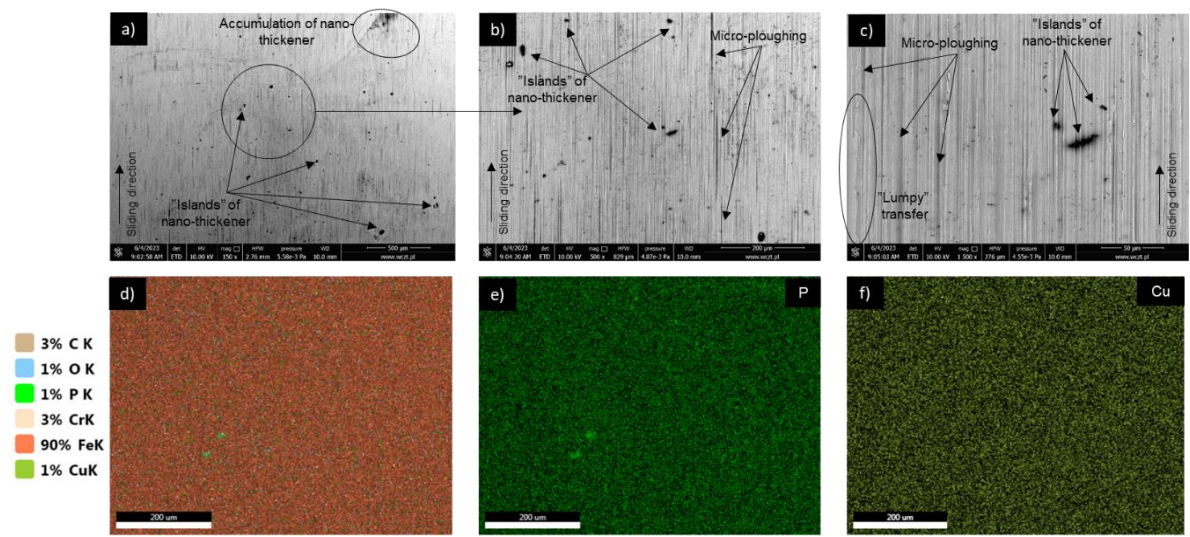

Figure 10. Investigations of wear signs on AISI 4130 rings (paired against UHMWPE) lubricated by IL + Cu-CNT0.1%: SEM pictures with different magnifications: 150x (a), 500x (b), 1500x (c), EDS analysis – general view (d), P content (e), Cu Content (f).

In turn, this may pose a problem for any tribo-active additives that will be applied during lubrication. In the contact zone, the lubricant encounters two different substrates with which it can react in a completely different way. As a result, there is a risk of improper formation of the boundary layer with varying strengths of bond with the substrate. In this case, the 'lumpy transfer' was not intense, which is confirmed only by small losses of material on the polymer surface and its slight transfer to the surface of the steel ring. Abrasive wear in the form of micro-ploughing is also identifiable on steel surfaces. The resulting grooves are primarily a consequence of the presence of micro-abrasives in the form of nano-carbon agglomerates formed during the frictional decomposition of the lubricant. The 'islands' of their clusters are visible in Fig. 10. The EDS analysis focused mainly on the presence of two elements with potential anti-friction and anti-wear properties, i.e. phosphorus and copper. The source of phosphorus is the selected IL ($C_{48}H_{102}O_4P_2$) here, which easily reacts with the steel substrate and forms tribo-active compounds with it [36]. As can be seen in Fig. 9d and 9e, P is concentrated mainly in the polymer roughness peaks, i.e. in the zone of direct contact with steel surfaces. In turn, Fig. 10d and 10e reveals the presence of P on steel surfaces. It is worth noting that, despite its small amount (~1%), it can be visible at single, concentration points, probably also at the roughness peaks. Perhaps the location of P is related to 'lumpy transfer' and filling the spaces free from the adhered polymer particles. A low content of Cu is observable on both the polymer and steel surfaces as can be seen in Figures 9f and 10f). However, the identification of such a low amount of an element using the EDS method may be questioned and difficult to distinguish from the so-called noise. Regardless, we cannot prove that the copper visible on polymer and steel surfaces comes from the lubricant used, its addition to CNTs had a beneficial effect on reducing friction and wear. For this reason, it is worth considering, at least hypothetically, the potential role of Cu in the friction of polymer–steel pairs. As a result of the decomposition of the lubricant structure during friction, agglomerates of copper nanoparticles are released from the CNT network and mix with the flowing polymer. After leaving contact, the polymer solidifies, and its surface is enriched with 'trapped' Cu nanoparticles. The presence of Cu on the steel surface is related to the phenomenon of its selective transfer in the presence of a low-density solvent (IL in this case) [44, 50-52]. Under such conditions, i.e. friction and dissolution, copper nanoparticles are released from the CNT network. Protected against oxidation by surrounding lubricant and very reactive, they easily bind with the steel surface, creating a very thin Cu nano-layer called servovitic film [44]. This layer is characterised by very low shear resistance, and consequently any adhesive joints that it forms with the counter-surfaces are broken along the original interface. Additionally, copper(II) ions, including ones formed at the surface termini via dangling OH bonds, enable the formation of partially soluble complexes, hence transferring copper compounds deeper and to the interfaces [46]. Therein, those compounds may be consequently reduced by the more electropositive iron phase, forming small copper nanoparticles [47].

Figures 11 and 12 present SEM pictures and EDS analysis of wear signs on the surfaces of POM-C blocks, and pair with them steel rings lubricated by IL + Cu-CNT0.1%.

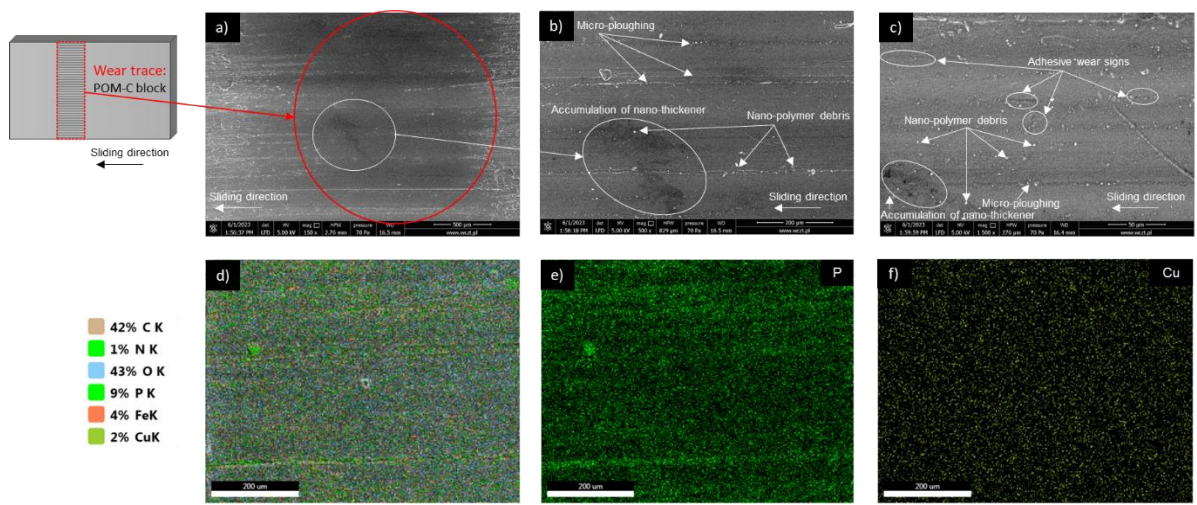

Figure 11. Investigations of wear signs on POM-C blocks lubricated by IL + Cu-CNT0.1%: SEM pictures with different magnification: 150x (a), 500x (b), 1500x (c), EDS analysis – general view (d), P content (e), Cu Content (f).

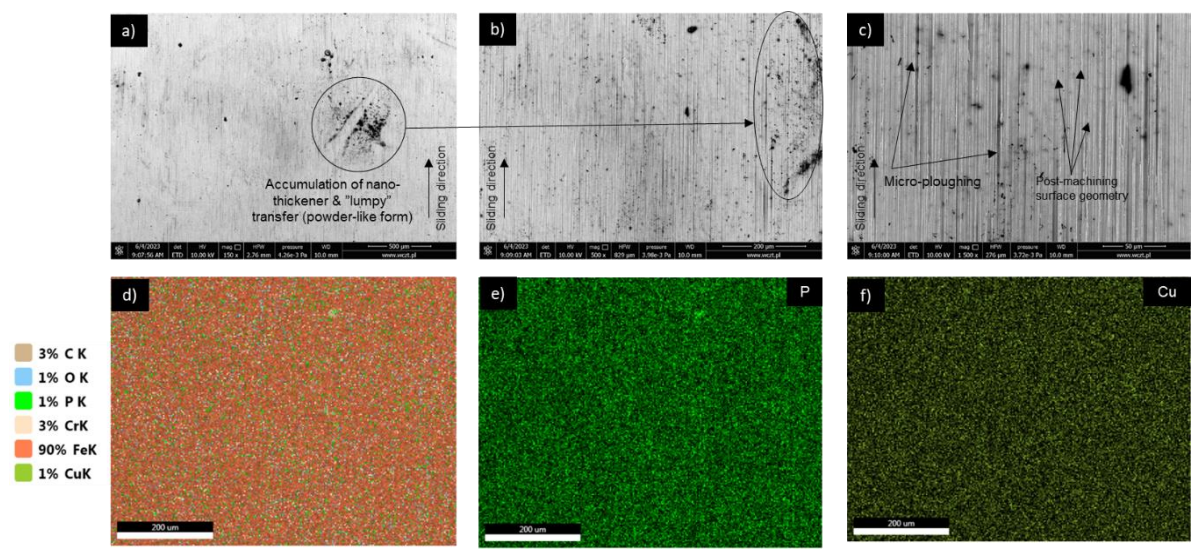

Figure 12. Investigations of wear signs on AISI 4130 rings (cooperated with POM-C) lubricated by IL + Cu-CNT0.1%: SEM pictures with different magnifications: 150x (a), 500x (b), 1500x (c), EDS analysis – general view (d), P content (e), Cu Content (f).

In the case of frictional interaction between POM-C and AISI 4130, the interpretation of wear traces resembles that for the first of the analysed polymers, with the dominant role of adhesive wear polymer and its 'lumpy transfer' to the steel surface. Interestingly, a slightly higher concentration of P on the polymer surface can be observed here. It can suggest a lower intensity of 'lumpy transfer' to the steel surface and can mean a higher share of protective layers of other origins (unfortunately, the concentration of P on steel ring surfaces does not confirm this theory). In turn, the higher Fe concentration within the wear mark of POM-C suggests that the steel surface could more intensively worn.

SEM pictures and EDS analysis for the last material configuration, PA–AISI 4130, are shown in Figures 13 and 14. In the case of polyamide polymer, the wear mechanism seems to be slightly different than for UHMWPE and POM-C. SEM images show signs of micro-ploughing as an effect of plastic deformations induced by the roughness peaks of the interacting steel surface. The size of the entire wear trace and individual grooves suggest that it was much more intense wear than the micro-adhesion identified for other polymers. Consequently, very clear signs of 'lumpy transfer' are visible on steel surfaces (Figure 14). This influenced the distribution of phosphorus on the polymer contact surface, the concentration of which is lower than for polyethylene and polyacetal. It can also be seen that the P concentration is consistent with the direction of sliding and the resulting post-abrasive grooves (Figure 13d and 13e).

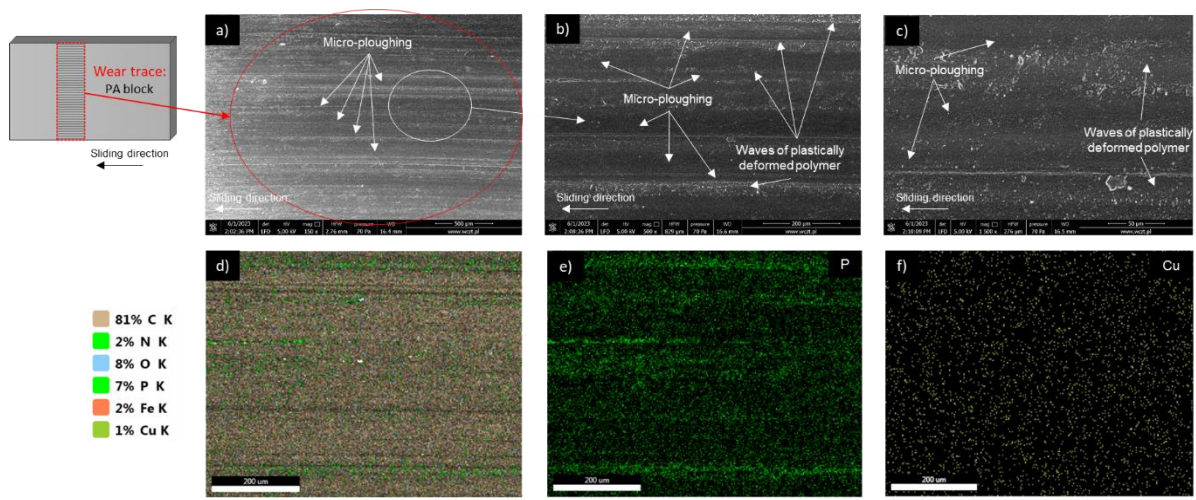

Figure 13. Investigations of wear signs on PA blocks lubricated by IL + Cu-CNT0.1%: SEM pictures with different magnifications: 150x (a), 500x (b), 1500x (c), EDS analysis – general view (d), P content (e), Cu Content (f).

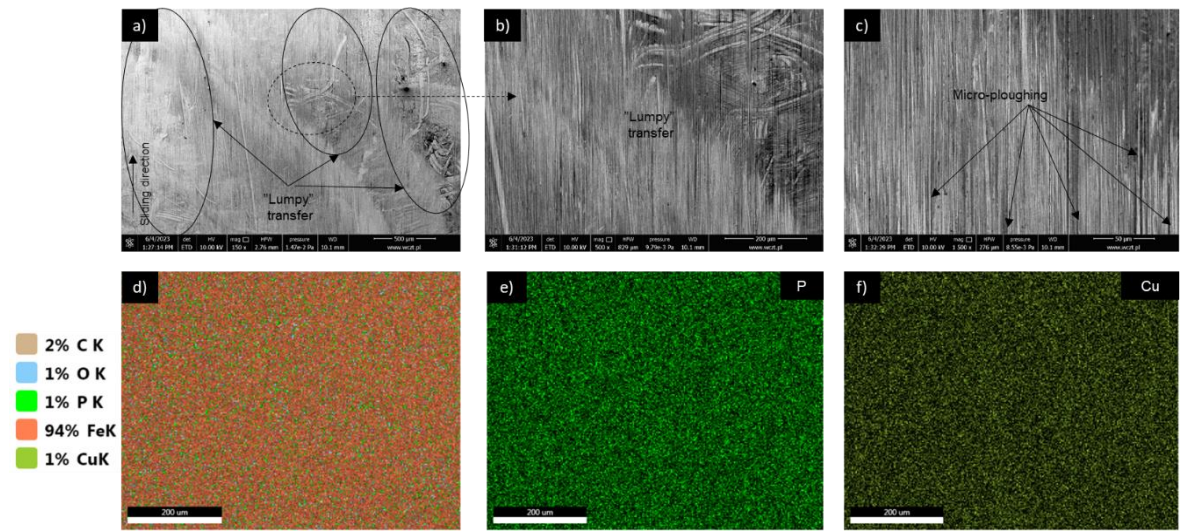

Figure 14. Investigations of wear signs on AISI 4130 rings (cooperated with PA) lubricated by IL + Cu-CNT0.1%: SEM pictures with different magnifications: 150x (a), 500x (b), 1500x (c), EDS analysis – general view (d), P content (e), Cu Content (f).

Imaging of wear traces on steel surfaces also indicates the presence of micro-ploughing, but they seem smaller than in the other cases (e.g. Figure 10c or Figure 12c). Perhaps this is related to the lack of large nanocarbon agglomerates that occurred in the AISI 4130 configurations with polyethylene and polyacetal.

Our interpretations of the wear mechanisms described above will be a starting point for a discussion on the protective effects that may result from IL + Cu-CNT0.1% lubrication (para 3.5).

### 3.3. Wear topography

The results of topographic measurements are auxiliary in these studies. This is due to the specific wear of polymers, characteristic of their interaction with harder materials, e.g. with steel. The plastic flow of polymers caused by friction means that the deformed material does not detach completely from the original surface. Despite a clear trace of wear, debris accumulates on its edges, and it is difficult to determine what reference surface should be considered as the correct one. Therefore, in this part, only selected examples of worn surfaces are presented, those that, in our opinion, best visualise the wear and the previously described mechanisms of its formation. Since the results of tribological tests for the configuration of friction pairs with polyamide seem to be the most promising – a topographic analysis was performed on the example of the wear of this polymer.

Figure 15 presents the 3D isometric views, Abbott-Firestone curves (AFC) and volumetric parameters of their characterisation for the new and worn surface of PA lubricated with the reference grease.

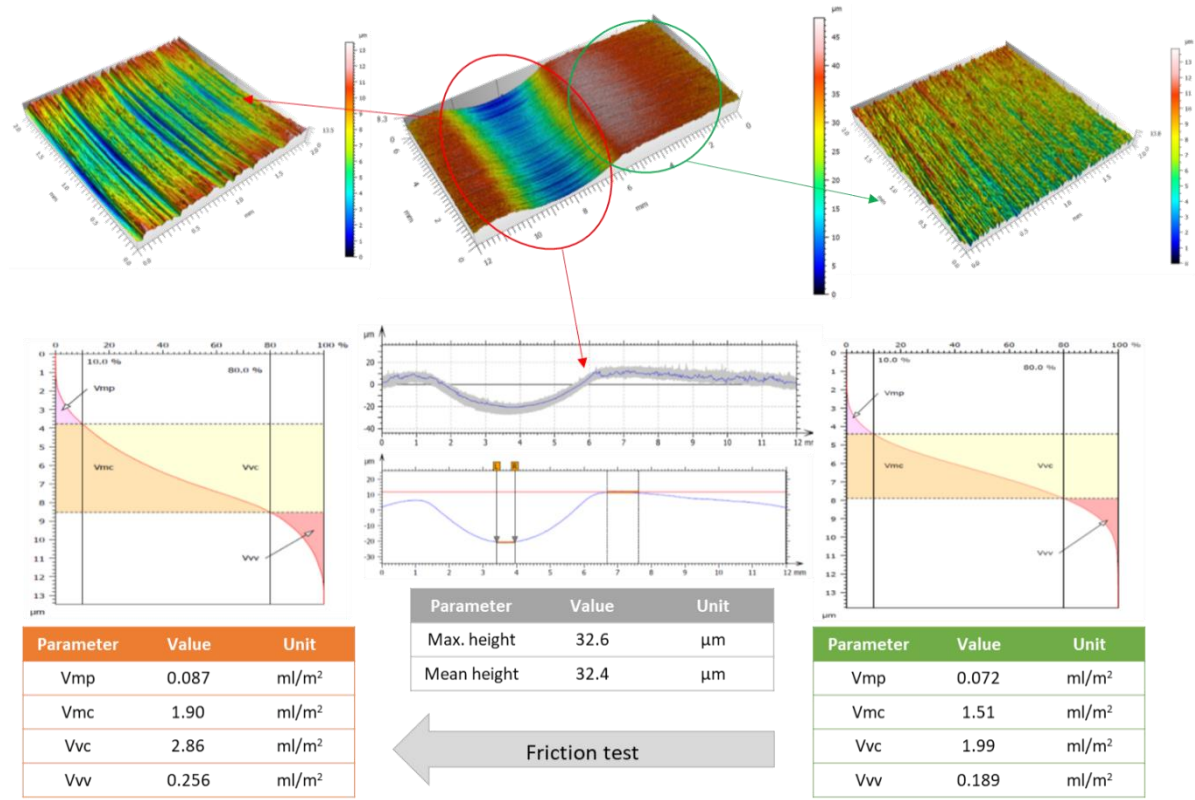

Figure 15. 3D isometric views, Abbott-Firestone curves, and volumetric parameters of their description for new (marked green) and worn (marked orange) surfaces of PA lubricated with the reference grease.

AFC analysis of the new polymer surface indicates the shape of the geometric structure typical for grinding, wherein (and this is less typical) the share of the valley part is more than twice as high (by volume) as the peak part (Vmp = 0.072 ml/m$^2$ vs Vvv = 0.189 ml/m$^2$). The same parameters measured after the friction test show a significant increase in values. What is characteristic is that the peak part increased only slightly, Vmp = 0.087 ml/m$^2$, and the valley part increased by almost 70 ml/m$^2$. This means the formation of many grooves characteristic of micro-ploughing and confirms that this wear mechanism accompanies 'lumpy transfer' during the friction of polyamide with steel (similar behaviour was found for PA lubricated with IL + Cu-CNT0.1%, see para 3.2 and Figure 13). The mean and maximum depth of the wear mark were also measured. The fact that these values are very close (~32.5 µm) may prove the absence of extreme grooves in the wear trace and its homogeneous nature. However, as mentioned previously, the reference point for measuring the depth of the wear trace remains uncertain. The edges of the trace may be 'lifted' owing to plastic deformation caused by the pressed ring, or may be enlarged by a wave of debris not detached from the original surface. In this case, considering these methodological weaknesses, the highest point of the average profile for the entire trace was used as the point against which depth was measured.

Figure 16 presents the 3D isometric views, Abbott-Firestone curves (AFC) and volumetric parameters of their characterisation for the new and worn surface of PA lubricated with the pure IL.

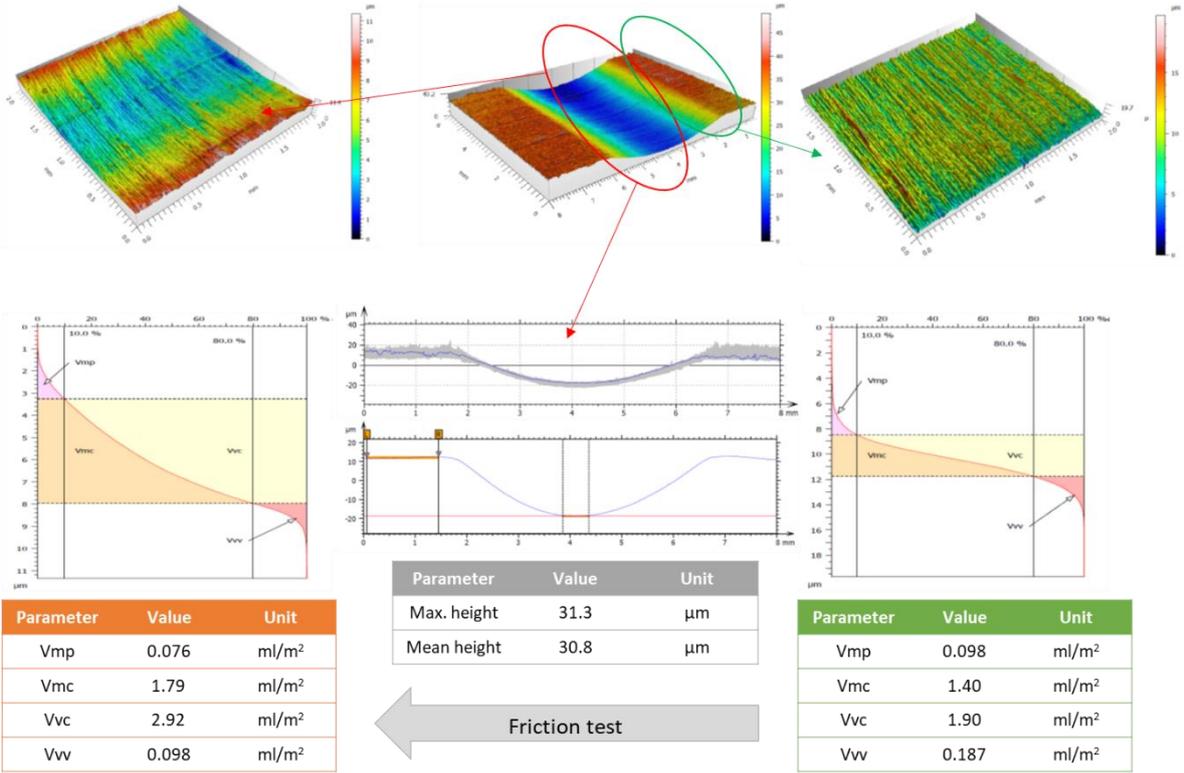

Figure 16. 3D isometric views, Abbott-Firestone curves, and volumetric parameters of their description for new (marked green) and worn (marked orange) surfaces of PA lubricated with the pure IL.

In general, a very similar evolution of roughness changes at the level of peaks and the core of the material can be observed as for lubrication with the reference grease. The main difference is the change in the volume of the material valleys: Vvv. In this case, you can notice the 'flattening' of this part of the surface geometry. This means that for lubrication with pure IL, we are dealing with plastic deformation of the polymer rather than its abrasive wear.

Figures 17–19 show the 3D isometric views, AFCs and volumetric parameters of their characterisation for the new and worn surfaces of PA lubricated with the IL thickened with different contents (wt) of CNTs (0.1, 0.5 and 1%).

In the first two cases (CNT 0.1% → Figure 17, CNT0.5% → Figure 18), it can be stated that a small addition of nano-thickener has an adverse effect on the depth of wear. For the lowest CNTs content used in the lubricant, the depth is almost twice as high as for the pure IL (and the reference grease). An increase in the content (wt) of CNTs in IL to 0.5% resulted in a decrease in the wear depth to values similar to, but still slightly higher (~ 46 µm) than for the reference grease. Only the highest content (wt) of CNTs in IL, i.e. 1%, caused a significant decrease in the wear depth to approximately 17.5 µm. Interestingly, these results did not coincide directly with the COF values in friction tests (Figure 8).

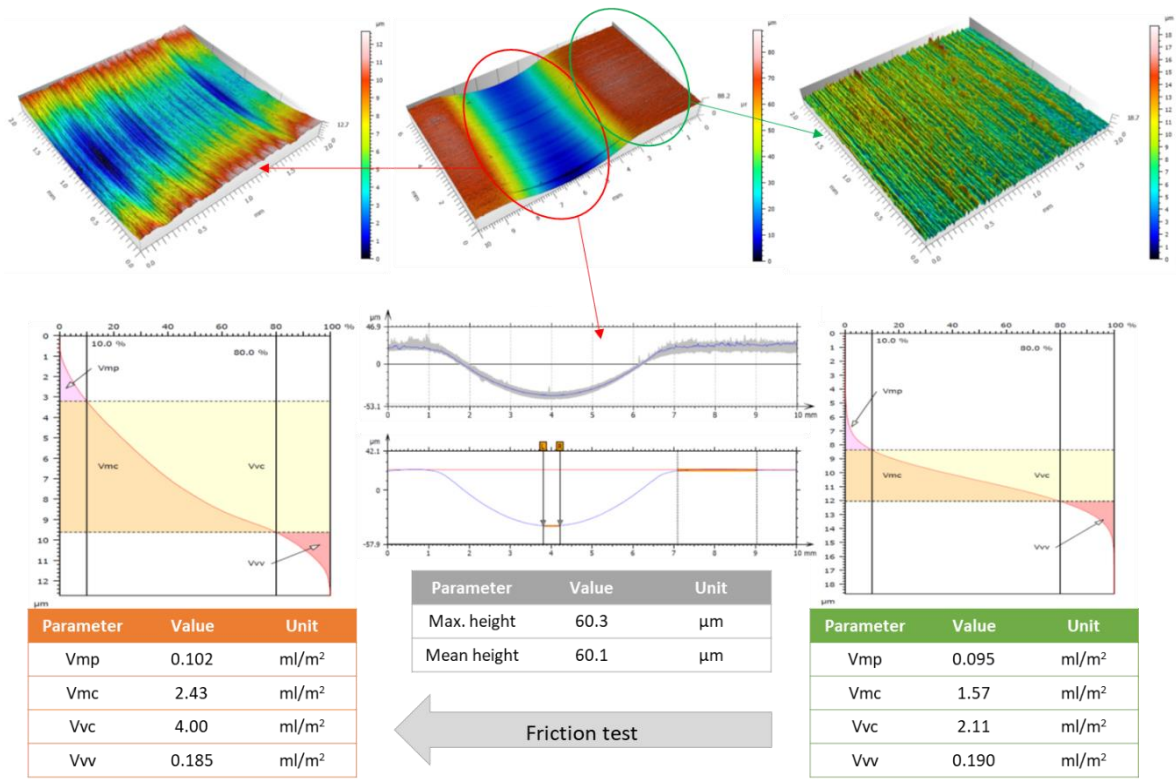

Figure 17. 3D isometric views, Abbott-Firestone curves and volumetric parameters (ISO25178) of their characterisation for the new (marked green) and worn (marked orange) surface of PA lubricated with the IL + CNT0.1%.

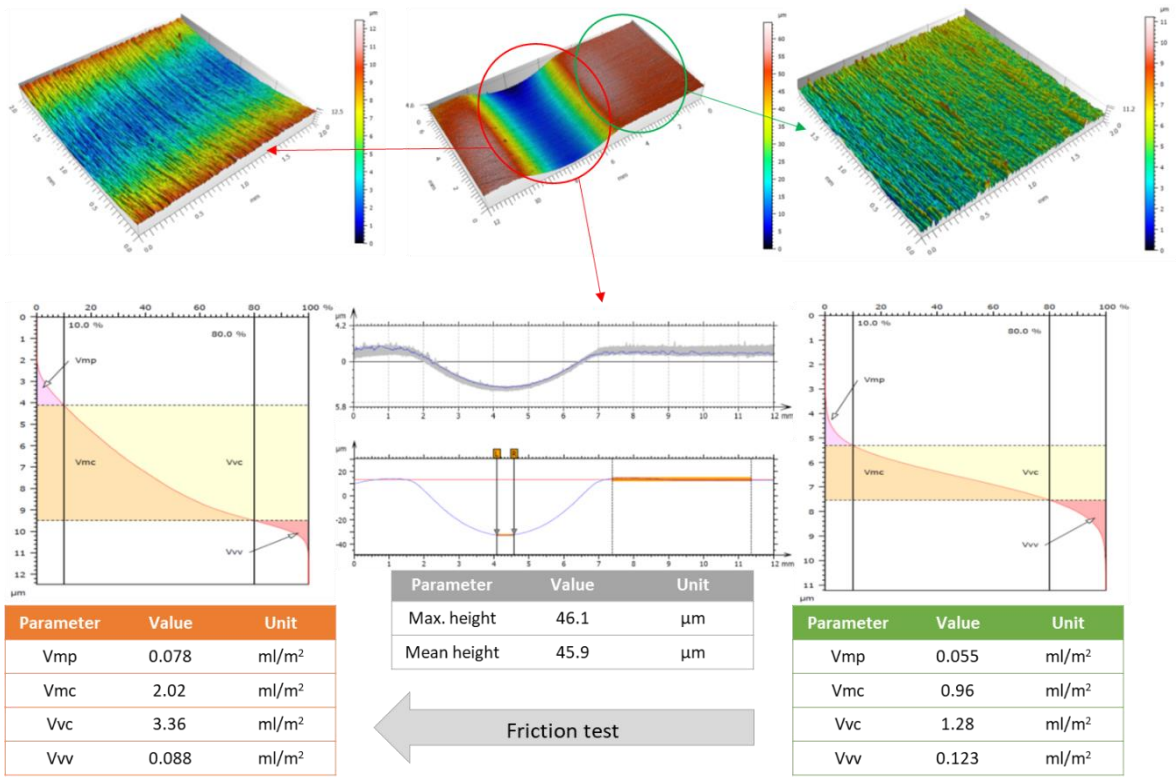

Figure 18. 3D isometric views, Abbott-Firestone curves and volumetric parameters (ISO25178) of their characterisation for the new (marked green) and worn (marked orange) surface of PA lubricated with the IL + CNT0.5%.

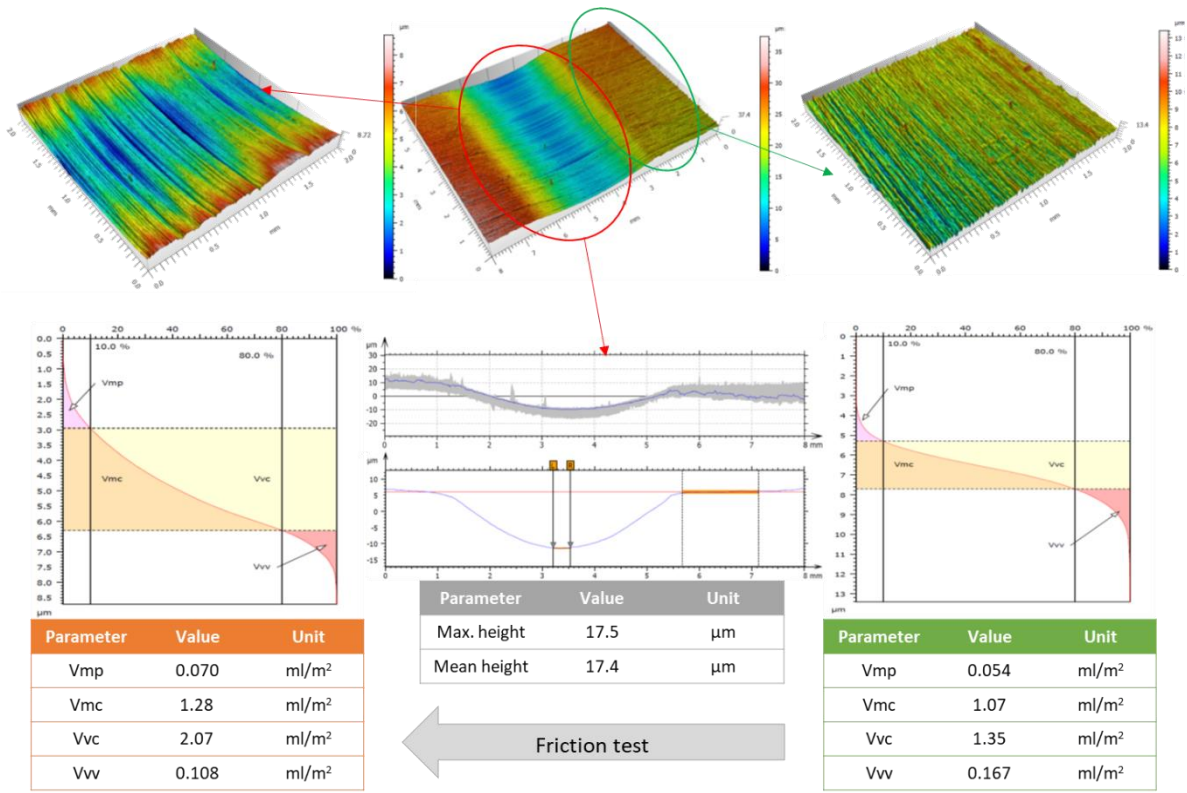

Figure 19. 3D isometric views, Abbott-Firestone curves and volumetric parameters (ISO25178) of their characterisation for the new (marked green) and worn (marked orange) surface of PA lubricated with the IL + CNT1%.

For the AISI4130–PA friction pair, it was found that the adding the highest amount of CNTs to IL has a beneficial effect on friction reduction. To explain this state of affairs, the analysis of AFC and its volumetric parameters is very helpful (because they describe all levels of the surface's geometric structure) [56-60]. Changes in the first parameter describing the volume of roughness peaks (Vmp) are ambiguous, that is why they will not be analysed further. In turn, in the case of parameters characterising the material core (Vmc and Vvc), we can observe a quite significant increase, when friction pair is lubricated by the IL with the addition of CNTs (0.5 and 0.1%). This can prove that owing to the pressure of the steel ring on the polymer, the surface was flattened, but also shallow micro-grooves appeared as a direct effect of the abrasive destruction mechanism. The additional decrease in valley volume (Vvv) makes this scenario more credible. Perhaps the lower amount of CNTs in the contact zone makes them more prone to agglomeration and generating abrasive rather than deformation resulting in exfoliation and the formation of nanocarbon slip planes needed to reduce friction [61-64]. The last analysed case in this part is the topography of the wear trace for a polyamide block lubricated with IL + Cu-CNT0.1%. 3D isometric views, Abbott-Firestone curves and volumetric parameters concerning this configuration are shown in Figure 20.

This configuration displays unexpected behaviour. On the one hand, we can observe the lowest depth of the wear trace (approx. 13 µm), which goes well with the positive assessment of IL + Cu-CNT0.1% as a lubricant with very good anti-friction properties. On the other hand, we can notice that the traces of wear, although flat, take the form of wide grooves resulting from plastic deformation rather than micro-ploughing. This is confirmed by the volumetric parameters, which have the highest values of all those analysed. To sum up, we are dealing here with a lubricant that provides the best friction conditions and the shallowest wear, but uncertainty about the contact area remains, which may be higher than in other cases.

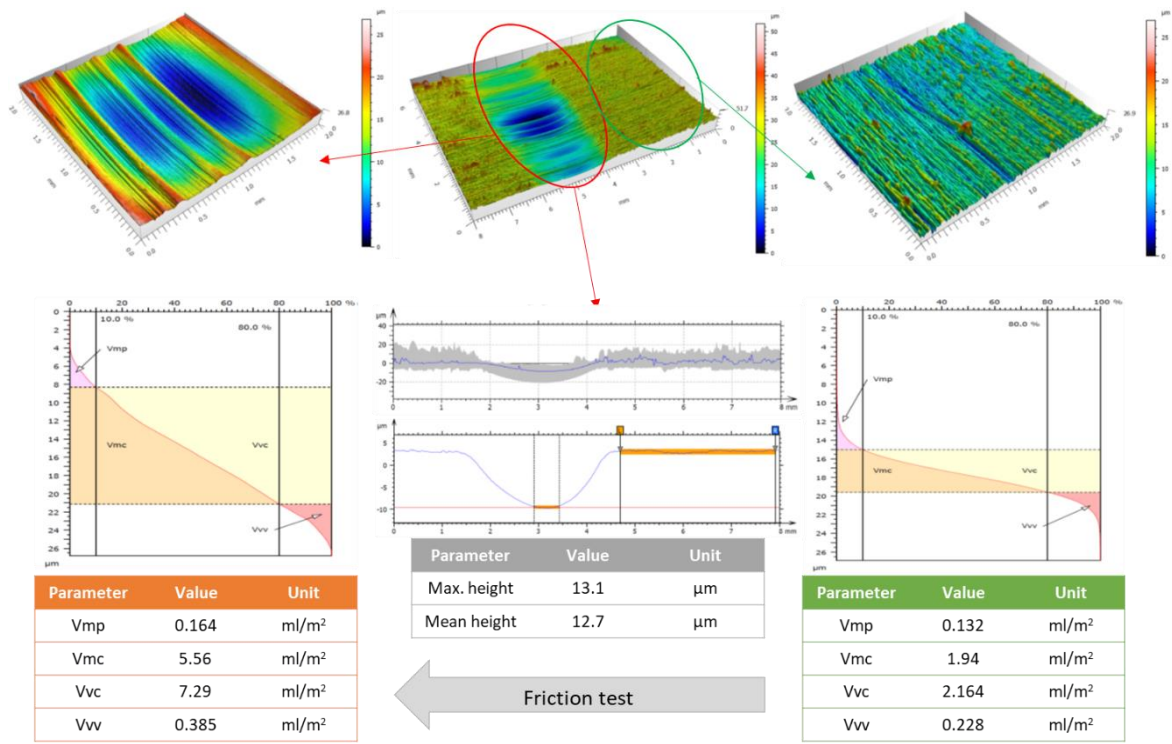

Figure 20. 3D isometric views, Abbott-Firestone curves and volumetric parameters (ISO25178) of their characterisation for the new (marked green) and worn (marked orange) surface of PA lubricated with the IL + Cu-CNT0.1%.

### 3.4. Surface free energy

The SFE consists of polar and dispersive components. The dispersive part of the solid–liquid interaction is related to electrostatic charge and is the dominant one in polymers. The polar interactions are related to dissymmetry in electron density in the molecules. As can be seen in Figure 21, the dispersive part of the new and unworn polymer surface displays a very high value, around 40 mN/m, with a very small polar component. The new metal surface has similar characteristics, with 28.6 mN/m for dispersive and 1.3 mN/m for polar components. In the metal/polymer contact, the transfer of the polymer onto the metallic surface quickly covers it with polymer film. Therefore, any analysis of the metal rings used in this study becomes redundant. The use of the reference grease lowered the SFE of the polymer, leaving the surface less active and prone to a higher wear rate. In all the configurations involving IL, there was a noticeable increase in the dispersive part and a significant increase in the polar component of all tested polymers. The increase in the overall SFE is mainly related to the IL, and the addition of CNTs or copper particles did not have any significant effect on the SFE. Higher free energy related to IL is most likely to result in better interactions of the proposed lubricants with newly created polymer surfaces during the wear and friction processes. Even a very small amount of lubricant will be easily adsorbed on the polymer surface, decreasing friction and protecting the interface.

### 3.5. Protective mechanisms

In paras 3.1-3.4, the friction behaviour, wear mechanisms and physicochemical changes (SFE) of polymers during frictional interaction with steel, under lubrication conditions with hybrid lubricants based on IL and CNTs, were characterised in detail. Both components of these innovative lubricants were found to have significant potential in reducing friction and wear. Therefore, the fundamental question arises: what protective effects do such lubricants generate on metal and polymer surfaces? In this part of the investigation, we propose a discussion of our views on this issue.

It appears that, for all material configurations, the primary polymer wear mechanism was 'lumpy transfer' caused by a combination of adhesive and abrasive wear. However, for different polymers the relationship between these wear mechanisms is not identical. For UHMWPE, adhesion plays the dominant role (Figure 9), while for POM-C and PA, a significant contribution of the abrasive component (ploughing) can be observed (Figures 11 and 13). These processes occur more easily when, as a result of frictionally increased temperature, the polymer begins to flow and its surface becomes more susceptible to adhesion and abrasion. At the same time, the lubricant in the contact zone can mix with the plasticised polymer, enriching its surface structure with tribo-active ingredients (tribo-active 'quasi-lakes'). When the polymer leaves frictional contact, its surface returns to a solid form and the lubricant components become trapped in the surface layer of the material. When re-entering the contact zone, before the polymer surface plasticises again, the lubricant components trapped there may be subjected to mechanical degradation. In the case of CNTs, this process may involve exfoliation of parts of the walls forming graphene-like flakes [61-64]. Under such conditions, friction and wear can be reduced dramatically. The idea of this process is shown in Figure 22.

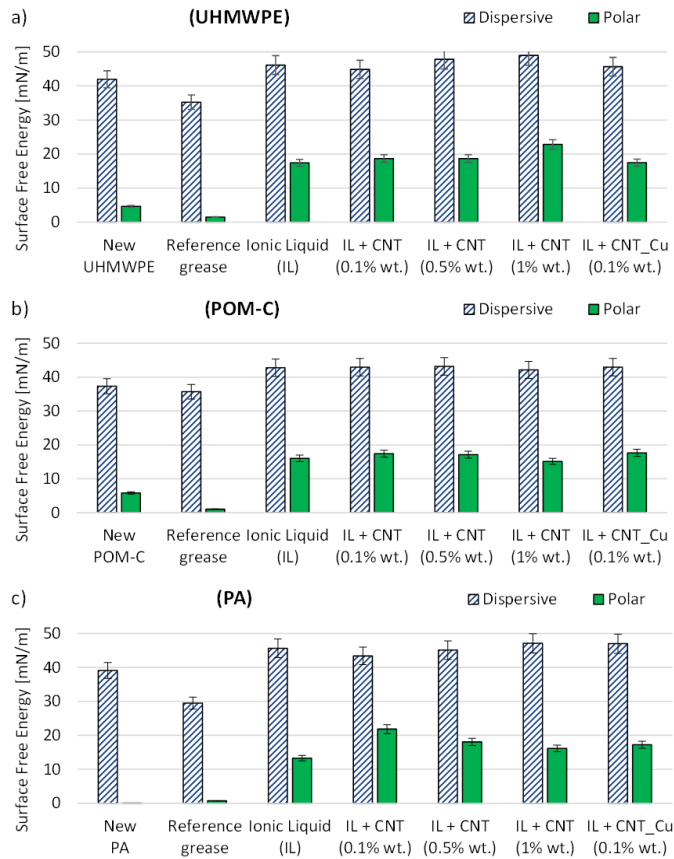

Figure 21. Changes in the SFE of polymers and an increase in the polar component related to the IL for the a) UHMWPE, b) POM-C and c) PA samples tested by wettability measurement of water and diiodomethane calculated using the OWRK method.

As mentioned earlier, the polymer protective layer formed due to 'lumpy transfer' on steel is heterogeneous. Therefore, what about those parts of the surface that do not have polymer lumps on them? In the case of lubrication by IL enriched only with CNTs, the adsorption of phosphorus on the steel surface seems to play a dominant role. However, it looks like intensive 'lumpy transfer' can significantly limit the effect of phosphorus in creating protective layers. When the roughness peaks of the steel surface are covered with polymer, the content of P in the contact zone can be limited. The mechanical and chemical decomposition of the oxide layer on the steel and the plastic flow of polymers due to friction cause both surfaces to be energetically stimulated and become an easy acceptor for phosphorus adsorption. This fact was confirmed by the EDS analysis in section. 3.2. The potential formation of the Cu-based servovitic films [44,50-52] on polymer and steel surfaces can be also conjectured. However, their formation was not confirmed in our research, therefore their presence on the surfaces of rubbing materials was left as a hypothesis and was not included in the interpretation of potential protective mechanisms (Figure 22).

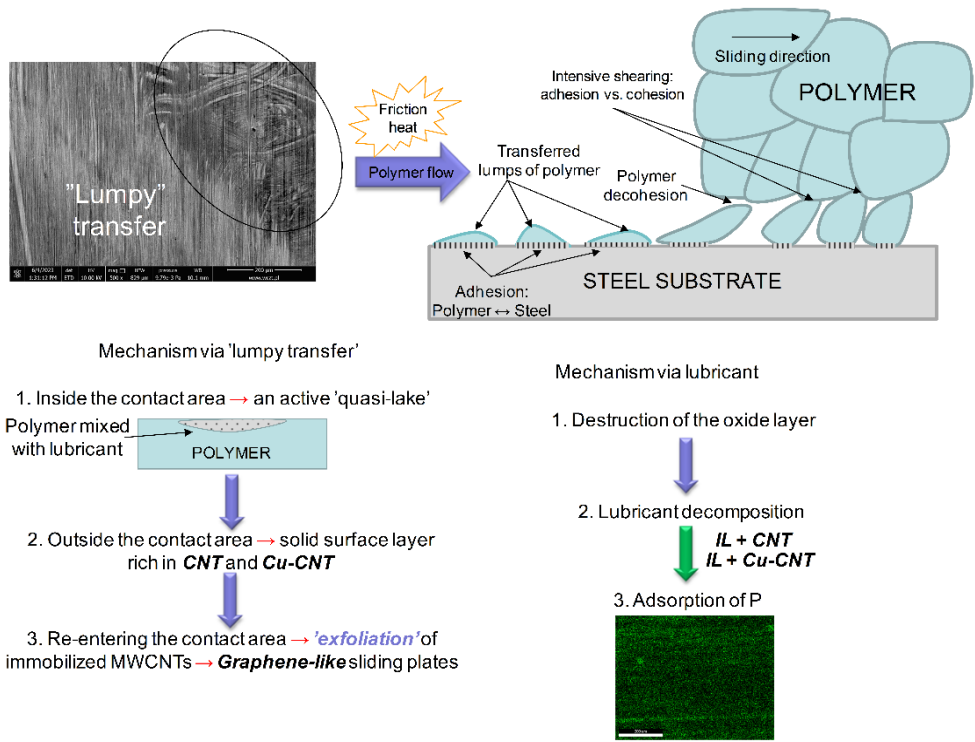

Figure 22. Protective mechanisms appearing in polymer–steel pairs lubricated with hybrid lubricants based on IL and CNTs.

Additionally, SFE measurements indicated that regardless of the content of the nanocarbon thickener, the presence of IL is the cause of the appearance of the polar component. It is this component that is responsible for the interaction with many lubricant additives. Therefore, it is possible that IL, acting as a light solvent, is simultaneously a donor of phosphorus atoms to the steel and polymer surfaces and a creator of conditions for the binding of other tribo-additives to the substrate. Such an addition may be CNTs which, when added in a sufficiently high amount to the applied IL, have a beneficial effect on reducing friction and wear in all analysed material configurations.

4. **Conclusions**

Based on the presented results, analysis and discussion, the following conclusions can be formulated:

- A new hybrid type of lubricant based on ionic liquid and carbon nanotubes as a thickener has been proposed. Concentrations of 0.1% to 1.0% wt of CNT formed stable dispersion of lubricants.
- A higher content of CNT (1% wt) in the lubricant favoured the reduction of friction and wear. This effect was obtained for all tested polymers: UHMWPE, POM-C and PA.
- The addition of the copper nanoparticles attached to the carbon nanotube mesh (as an additive to IL at a concentration of 0.1%) allowed for the highest reduction in friction for UHMWPE–AISI4130 and POM-C–AISI4130 pairs and a significant reduction in wear (measured by trace depth) for all polymers.
- The adsorption of phosphorus onto polymer surfaces has been observed and its accumulation is located on the roughness peaks and/or related to the sliding direction. Only rare, single points of P concentration are identifiable on steel surfaces.
- A 'lumpy transfer' is the dominant form of wear for all friction test configurations, with signs of abrasive and adhesive wear also taking place.
- The presence of ionic liquid in contact induced an increase in the polar component of SFE on polymer surfaces. This is a factor that may contribute to the formation of a protective film on rubbing surfaces.

## 5. Perspectives

The addition of copper nanoparticles to CNTs has a clear influence on friction and wear in steel–polymer pairs. Therefore, other types of nanoparticles doping to CNTs can be explored further. Also, the type of interactions and the form that CNTs can take once in contact and under stress (nanotubes, nano-rolls or nano-flakes) should be analysed in more detail to explore other possible protective mechanisms.

## Acknowledgement

The authors would like to acknowledge the financial support for this research provided by the National Science Centre in Poland (Project 2020/39/B/ST5/02562). In addition, K.J. Kubiak would like to acknowledge the support provided by the UK EPSRC grant TRENT (EP/S030476/1).

# Appendix A – Supplementary data

Table S1. Results of friction torque measurements in 60s intervals, their average values, and standard deviations for friction pairs with the UHMWPE component.

| Polymer: Tivar 1000 (UHMWPE); Lubricant: $C_{48}H_{102}O_4P_2$ | | | | | | Polymer: Tivar 1000 (UHMWPE); Lubricant: $C_{48}H_{102}O_4P_2$+CNT1% | | | | | |
|---|---|---|---|---|---|---|---|---|---|---|---|
| Time [s] | Friction torque [Nm] | | | | | Time [s] | Friction torque [Nm] | | | | |
| | Test 1 | Test 2 | Test 3 | Average | St. Dev. | | Test 1 | Test 2 | Test 3 | Average | St. Dev. |
| 60 | 0.2 | 0.2 | 0.2 | 0.20 | 0.00 | 60 | 0.15 | 0.15 | 0.15 | 0.15 | 0.00 |
| 120 | 0.4 | 0.5 | 0.4 | 0.43 | 0.05 | 120 | 0.45 | 0.55 | 0.45 | 0.48 | 0.05 |
| 180 | 0.7 | 0.85 | 0.8 | 0.78 | 0.06 | 180 | 0.75 | 0.85 | 0.65 | 0.75 | 0.08 |
| 240 | 0.7 | 0.85 | 0.8 | 0.78 | 0.06 | 240 | 0.8 | 0.85 | 0.65 | 0.77 | 0.08 |
| 300 | 0.7 | 0.9 | 0.8 | 0.80 | 0.08 | 300 | 0.8 | 0.85 | 0.65 | 0.77 | 0.08 |
| 360 | 0.7 | 0.9 | 0.8 | 0.80 | 0.08 | 360 | 0.75 | 0.85 | 0.65 | 0.75 | 0.08 |
| 420 | 0.7 | 0.9 | 0.8 | 0.80 | 0.08 | 420 | 0.75 | 0.85 | 0.65 | 0.75 | 0.08 |
| 480 | 0.75 | 0.9 | 0.9 | 0.85 | 0.07 | 480 | 0.75 | 0.85 | 0.6 | 0.73 | 0.10 |
| 540 | 0.75 | 0.9 | 0.9 | 0.85 | 0.07 | 540 | 0.75 | 0.8 | 0.6 | 0.72 | 0.08 |
| 600 | 0.75 | 0.9 | 0.9 | 0.85 | 0.07 | 600 | 0.75 | 0.8 | 0.6 | 0.72 | 0.08 |
| 660 | 0.75 | 0.9 | 0.95 | 0.87 | 0.08 | 660 | 0.75 | 0.8 | 0.6 | 0.72 | 0.08 |
| 720 | 0.7 | 0.9 | 0.95 | 0.85 | 0.11 | 720 | 0.7 | 0.75 | 0.55 | 0.67 | 0.08 |
| 780 | 0.7 | 0.9 | 0.95 | 0.85 | 0.11 | 780 | 0.7 | 0.75 | 0.55 | 0.67 | 0.08 |
| 840 | 0.7 | 0.9 | 0.95 | 0.85 | 0.11 | 840 | 0.65 | 0.7 | 0.55 | 0.63 | 0.06 |
| 900 | 0.65 | 0.9 | 0.95 | 0.83 | 0.13 | 900 | 0.65 | 0.7 | 0.5 | 0.62 | 0.08 |
| 960 | 0.65 | 0.9 | 0.95 | 0.83 | 0.13 | 960 | 0.65 | 0.7 | 0.5 | 0.62 | 0.08 |
| 1020 | 0.65 | 0.9 | 1 | 0.85 | 0.15 | 1020 | 0.65 | 0.7 | 0.5 | 0.62 | 0.08 |
| 1080 | 0.65 | 0.95 | 1 | 0.87 | 0.15 | 1080 | 0.7 | 0.65 | 0.45 | 0.60 | 0.11 |
| 1140 | 0.65 | 0.95 | 1 | 0.87 | 0.15 | 1140 | 0.65 | 0.65 | 0.45 | 0.58 | 0.09 |
| 1200 | 0.65 | 0.95 | 1 | 0.87 | 0.15 | 1200 | 0.65 | 0.65 | 0.45 | 0.58 | 0.09 |
| 1260 | 0.65 | 0.95 | 1 | 0.87 | 0.15 | 1260 | 0.65 | 0.65 | 0.45 | 0.58 | 0.09 |
| 1320 | 0.65 | 0.95 | 1 | 0.87 | 0.15 | 1320 | 0.65 | 0.65 | 0.45 | 0.58 | 0.09 |
| 1380 | 0.65 | 0.95 | 1 | 0.87 | 0.15 | 1380 | 0.65 | 0.65 | 0.45 | 0.58 | 0.09 |
| 1440 | 0.65 | 0.95 | 1 | 0.87 | 0.15 | 1440 | 0.65 | 0.65 | 0.45 | 0.58 | 0.09 |
| 1500 | 0.65 | 0.9 | 1.05 | 0.87 | 0.16 | 1500 | 0.65 | 0.65 | 0.45 | 0.58 | 0.09 |
| 1560 | 0.65 | 0.9 | 1.05 | 0.87 | 0.16 | 1560 | 0.65 | 0.65 | 0.45 | 0.58 | 0.09 |
| 1620 | 0.65 | 0.9 | 1.05 | 0.87 | 0.16 | 1620 | 0.65 | 0.65 | 0.45 | 0.58 | 0.09 |
| 1680 | 0.65 | 0.95 | 1.05 | 0.88 | 0.17 | 1680 | 0.65 | 0.65 | 0.45 | 0.58 | 0.09 |
| 1740 | 0.65 | 0.95 | 1.05 | 0.88 | 0.17 | 1740 | 0.65 | 0.65 | 0.45 | 0.58 | 0.09 |
| 1800 | 0.65 | 0.95 | 1.05 | 0.88 | 0.17 | 1800 | 0.65 | 0.65 | 0.45 | 0.58 | 0.09 |
| Polymer: Tivar 1000 (UHMWPE); Lubricant: $C_{48}H_{102}O_4P_2$+CNT0.5% | | | | | | Polymer: Tivar 1000 (UHMWPE); Lubricant: $C_{48}H_{102}O_4P_2$+CNT0.1% | | | | | |
| Time [s] | Friction torque [Nm] | | | | | Time [s] | Friction torque [Nm] | | | | |
| | Test 1 | Test 2 | Test 3 | Average | St. Dev. | | Test 1 | Test 2 | Test 3 | Average | St. Dev. |
| 60 | 0.15 | 0.15 | 0.15 | 0.15 | 0.00 | 60 | 0.1 | 0.15 | 0.15 | 0.13 | 0.02 |
| 120 | 0.45 | 0.55 | 0.45 | 0.48 | 0.05 | 120 | 0.45 | 0.5 | 0.45 | 0.47 | 0.02 |
| 180 | 0.8 | 0.9 | 0.7 | 0.80 | 0.08 | 180 | 0.65 | 0.8 | 0.75 | 0.73 | 0.06 |
| 240 | 0.8 | 0.9 | 0.8 | 0.83 | 0.05 | 240 | 0.65 | 0.8 | 0.75 | 0.73 | 0.06 |
| 300 | 0.8 | 0.9 | 0.8 | 0.83 | 0.05 | 300 | 0.7 | 0.8 | 0.75 | 0.75 | 0.04 |
| 360 | 0.8 | 1 | 0.7 | 0.83 | 0.12 | 360 | 0.65 | 0.85 | 0.9 | 0.80 | 0.11 |
| 420 | 0.8 | 1 | 0.7 | 0.83 | 0.12 | 420 | 0.65 | 0.85 | 0.9 | 0.80 | 0.11 |
| 480 | 0.8 | 1 | 0.6 | 0.80 | 0.16 | 480 | 0.65 | 0.9 | 1.1 | 0.88 | 0.18 |
| 540 | 0.9 | 1 | 0.65 | 0.85 | 0.15 | 540 | 0.65 | 0.9 | 1.1 | 0.88 | 0.18 |
| 600 | 0.9 | 1 | 0.65 | 0.85 | 0.15 | 600 | 0.65 | 0.9 | 1.1 | 0.88 | 0.18 |
| 660 | 0.9 | 1 | 0.8 | 0.90 | 0.08 | 660 | 0.65 | 0.9 | 1.1 | 0.88 | 0.18 |
| 720 | 0.85 | 1 | 0.8 | 0.88 | 0.08 | 720 | 0.65 | 1 | 1 | 0.88 | 0.16 |
| 780 | 0.85 | 1 | 0.9 | 0.92 | 0.06 | 780 | 0.65 | 1 | 1 | 0.88 | 0.16 |
| 840 | 0.85 | 1 | 0.9 | 0.92 | 0.06 | 840 | 0.65 | 1 | 1 | 0.88 | 0.16 |
| 900 | 0.9 | 1 | 0.9 | 0.93 | 0.05 | 900 | 0.6 | 1 | 0.9 | 0.83 | 0.17 |
| 960 | 0.9 | 1 | 0.9 | 0.93 | 0.05 | 960 | 0.6 | 1 | 0.8 | 0.80 | 0.16 |
| 1020 | 0.9 | 1 | 0.9 | 0.93 | 0.05 | 1020 | 0.6 | 1 | 0.8 | 0.80 | 0.16 |
| 1080 | 0.9 | 1 | 0.9 | 0.93 | 0.05 | 1080 | 0.6 | 1 | 0.7 | 0.77 | 0.17 |
| 1140 | 0.9 | 1 | 0.9 | 0.93 | 0.05 | 1140 | 0.6 | 1 | 0.9 | 0.83 | 0.17 |
| 1200 | 0.9 | 1 | 0.8 | 0.90 | 0.08 | 1200 | 0.6 | 1 | 1 | 0.87 | 0.19 |
| 1260 | 0.9 | 1 | 0.8 | 0.90 | 0.08 | 1260 | 0.6 | 1 | 1 | 0.87 | 0.19 |
| 1320 | 0.9 | 1 | 0.8 | 0.90 | 0.08 | 1320 | 0.6 | 1.05 | 1 | 0.88 | 0.20 |
| 1380 | 0.95 | 1 | 0.8 | 0.92 | 0.08 | 1380 | 0.6 | 1.05 | 0.8 | 0.82 | 0.18 |
| 1440 | 0.95 | 1 | 0.8 | 0.92 | 0.08 | 1440 | 0.6 | 1.05 | 0.8 | 0.82 | 0.18 |
| 1500 | 0.95 | 1 | 0.8 | 0.92 | 0.08 | 1500 | 0.6 | 1.05 | 0.9 | 0.85 | 0.19 |
| 1560 | 0.95 | 1.05 | 0.9 | 0.97 | 0.06 | 1560 | 0.6 | 1.05 | 1 | 0.88 | 0.20 |
| 1620 | 0.95 | 1.05 | 1 | 1.00 | 0.04 | 1620 | 0.6 | 1.05 | 1 | 0.88 | 0.20 |
| 1680 | 0.9 | 1.05 | 1 | 0.98 | 0.06 | 1680 | 0.6 | 1.05 | 1 | 0.88 | 0.20 |
| 1740 | 0.9 | 1.05 | 1.1 | 1.02 | 0.08 | 1740 | 0.6 | 1.05 | 1 | 0.88 | 0.20 |
| 1800 | 0.9 | 1.05 | 1.1 | 1.02 | 0.08 | 1800 | 0.6 | 1.05 | 1 | 0.88 | 0.20 |

Table S1. Results of friction torque measurements in 60s intervals, their average values, and standard deviations for friction pairs with the UHMWPE component (continued).

| Polymer: Tivar 1000 (UHMWPE); Lubricant: $C_{48}H_{102}O_4P_2$+Cu_CNT0.1% | | | | | | Polymer: Tivar 1000 (UHMWPE); Lubricant: Reference | | | | | |
|---|---|---|---|---|---|---|---|---|---|---|---|
| Time [s] | Friction torque [Nm] | | | | | Time [s] | Friction torque [Nm] | | | | |
| | Test 1 | Test 2 | Test 3 | Average | St. Dev. | | Test 1 | Test 2 | Test 3 | Average | St. Dev. |
| 60 | 0.1 | 0.15 | 0.15 | 0.13 | 0.02 | 60 | 0.2 | 0.15 | 0.15 | 0.17 | 0.02 |
| 120 | 0.3 | 0.45 | 0.35 | 0.37 | 0.06 | 120 | 0.5 | 0.35 | 0.3 | 0.38 | 0.08 |
| 180 | 0.4 | 0.65 | 0.35 | 0.47 | 0.13 | 180 | 0.7 | 0.55 | 0.5 | 0.58 | 0.08 |
| 240 | 0.4 | 0.65 | 0.35 | 0.47 | 0.13 | 240 | 0.7 | 0.55 | 0.5 | 0.58 | 0.08 |
| 300 | 0.5 | 0.65 | 0.35 | 0.50 | 0.12 | 300 | 0.75 | 0.75 | 0.45 | 0.65 | 0.14 |
| 360 | 0.5 | 0.65 | 0.35 | 0.50 | 0.12 | 360 | 0.75 | 0.85 | 0.45 | 0.68 | 0.17 |
| 420 | 0.5 | 0.75 | 0.35 | 0.53 | 0.16 | 420 | 0.75 | 0.85 | 0.45 | 0.68 | 0.17 |
| 480 | 0.5 | 0.75 | 0.35 | 0.53 | 0.16 | 480 | 0.75 | 0.75 | 0.45 | 0.65 | 0.14 |
| 540 | 0.5 | 0.75 | 0.35 | 0.53 | 0.16 | 540 | 0.75 | 0.75 | 0.45 | 0.65 | 0.14 |
| 600 | 0.5 | 0.75 | 0.35 | 0.53 | 0.16 | 600 | 0.75 | 0.75 | 0.45 | 0.65 | 0.14 |
| 660 | 0.5 | 0.75 | 0.35 | 0.53 | 0.16 | 660 | 0.75 | 0.75 | 0.45 | 0.65 | 0.14 |
| 720 | 0.45 | 0.65 | 0.35 | 0.48 | 0.12 | 720 | 0.75 | 0.7 | 0.55 | 0.67 | 0.08 |
| 780 | 0.45 | 0.65 | 0.35 | 0.48 | 0.12 | 780 | 0.75 | 0.7 | 0.55 | 0.67 | 0.08 |
| 840 | 0.45 | 0.65 | 0.35 | 0.48 | 0.12 | 840 | 0.9 | 0.7 | 0.75 | 0.78 | 0.08 |
| 900 | 0.45 | 0.65 | 0.35 | 0.48 | 0.12 | 900 | 0.9 | 0.8 | 0.75 | 0.82 | 0.06 |
| 960 | 0.45 | 0.65 | 0.35 | 0.48 | 0.12 | 960 | 0.9 | 0.8 | 0.9 | 0.87 | 0.05 |
| 1020 | 0.45 | 0.65 | 0.35 | 0.48 | 0.12 | 1020 | 0.9 | 0.8 | 1 | 0.90 | 0.08 |
| 1080 | 0.45 | 0.65 | 0.35 | 0.48 | 0.12 | 1080 | 0.9 | 0.8 | 1 | 0.90 | 0.08 |
| 1140 | 0.45 | 0.65 | 0.35 | 0.48 | 0.12 | 1140 | 0.9 | 0.8 | 1 | 0.90 | 0.08 |
| 1200 | 0.45 | 0.65 | 0.35 | 0.48 | 0.12 | 1200 | 0.8 | 0.65 | 1 | 0.82 | 0.14 |
| 1260 | 0.45 | 0.65 | 0.35 | 0.48 | 0.12 | 1260 | 0.8 | 0.65 | 1 | 0.82 | 0.14 |
| 1320 | 0.4 | 0.65 | 0.35 | 0.47 | 0.13 | 1320 | 0.8 | 0.65 | 1 | 0.82 | 0.14 |
| 1380 | 0.4 | 0.65 | 0.35 | 0.47 | 0.13 | 1380 | 0.8 | 0.65 | 1 | 0.82 | 0.14 |
| 1440 | 0.4 | 0.65 | 0.35 | 0.47 | 0.13 | 1440 | 0.8 | 0.65 | 1 | 0.82 | 0.14 |
| 1500 | 0.4 | 0.65 | 0.35 | 0.47 | 0.13 | 1500 | 0.9 | 0.65 | 1 | 0.85 | 0.15 |
| 1560 | 0.4 | 0.65 | 0.35 | 0.47 | 0.13 | 1560 | 0.9 | 0.65 | 1 | 0.85 | 0.15 |
| 1620 | 0.4 | 0.65 | 0.35 | 0.47 | 0.13 | 1620 | 0.9 | 0.65 | 1 | 0.85 | 0.15 |
| 1680 | 0.4 | 0.65 | 0.35 | 0.47 | 0.13 | 1680 | 0.9 | 0.65 | 1.1 | 0.88 | 0.18 |
| 1740 | 0.4 | 0.65 | 0.35 | 0.47 | 0.13 | 1740 | 0.9 | 0.65 | 1.1 | 0.88 | 0.18 |
| 1800 | 0.4 | 0.65 | 0.35 | 0.47 | 0.13 | 1800 | 1 | 0.65 | 1.1 | 0.92 | 0.19 |

Table S2. Results of friction torque measurements in 60s intervals, their average values, and standard deviations for friction pairs with the POM-C component.

| Polymer: Ertacetal C (POM-C); Lubricant: $C_{48}H_{102}O_4P_2$ | | | | | | Polymer: Ertacetal C (POM-C); Lubricant: $C_{48}H_{102}O_4P_2$+CNT1% | | | | | |
|---|---|---|---|---|---|---|---|---|---|---|---|
| Time [s] | Friction torque [Nm] | | | | | Time [s] | Friction torque [Nm] | | | | |
| | Test 1 | Test 2 | Test 3 | Average | St. Dev. | | Test 1 | Test 2 | Test 3 | Average | St. Dev. |
| 60 | 0.35 | 0.3 | 0.35 | 0.33 | 0.02 | 60 | 0.2 | 0.25 | 0.2 | 0.22 | 0.02 |
| 120 | 0.7 | 0.6 | 0.7 | 0.67 | 0.05 | 120 | 0.45 | 0.55 | 0.5 | 0.50 | 0.04 |
| 180 | 0.95 | 0.75 | 0.95 | 0.88 | 0.09 | 180 | 0.7 | 0.85 | 0.8 | 0.78 | 0.06 |
| 240 | 0.95 | 0.75 | 0.95 | 0.88 | 0.09 | 240 | 0.7 | 0.85 | 0.8 | 0.78 | 0.06 |
| 300 | 0.95 | 0.75 | 0.95 | 0.88 | 0.09 | 300 | 0.7 | 0.8 | 0.8 | 0.77 | 0.05 |
| 360 | 0.95 | 0.75 | 0.9 | 0.87 | 0.08 | 360 | 0.65 | 0.8 | 0.8 | 0.75 | 0.07 |
| 420 | 0.95 | 0.75 | 0.9 | 0.87 | 0.08 | 420 | 0.65 | 0.8 | 0.8 | 0.75 | 0.07 |
| 480 | 0.95 | 0.75 | 0.85 | 0.85 | 0.08 | 480 | 0.65 | 0.8 | 0.8 | 0.75 | 0.07 |
| 540 | 0.95 | 0.75 | 0.85 | 0.85 | 0.08 | 540 | 0.6 | 0.8 | 0.8 | 0.73 | 0.09 |
| 600 | 0.9 | 0.75 | 0.85 | 0.83 | 0.06 | 600 | 0.6 | 0.8 | 0.8 | 0.73 | 0.09 |
| 660 | 0.9 | 0.75 | 0.85 | 0.83 | 0.06 | 660 | 0.6 | 0.95 | 0.8 | 0.78 | 0.14 |
| 720 | 0.9 | 0.75 | 0.85 | 0.83 | 0.06 | 720 | 0.6 | 0.95 | 0.8 | 0.78 | 0.14 |
| 780 | 0.9 | 0.75 | 0.85 | 0.83 | 0.06 | 780 | 0.6 | 0.95 | 0.8 | 0.78 | 0.14 |
| 840 | 0.85 | 0.75 | 0.8 | 0.80 | 0.04 | 840 | 0.6 | 1.1 | 0.75 | 0.82 | 0.21 |
| 900 | 0.85 | 0.7 | 0.8 | 0.78 | 0.06 | 900 | 0.6 | 1.2 | 0.75 | 0.85 | 0.25 |
| 960 | 0.85 | 0.7 | 0.8 | 0.78 | 0.06 | 960 | 0.55 | 1.3 | 0.75 | 0.87 | 0.32 |
| 1020 | 0.85 | 0.65 | 0.8 | 0.77 | 0.08 | 1020 | 0.55 | 1.35 | 0.75 | 0.88 | 0.34 |
| 1080 | 0.85 | 0.65 | 0.75 | 0.75 | 0.08 | 1080 | 0.55 | 1.35 | 0.75 | 0.88 | 0.34 |
| 1140 | 0.85 | 0.65 | 0.75 | 0.75 | 0.08 | 1140 | 0.55 | 1.45 | 0.75 | 0.92 | 0.39 |
| 1200 | 0.85 | 0.65 | 0.75 | 0.75 | 0.08 | 1200 | 0.55 | 1.45 | 0.7 | 0.90 | 0.39 |
| 1260 | 0.85 | 0.65 | 0.75 | 0.75 | 0.08 | 1260 | 0.55 | 1.4 | 0.7 | 0.88 | 0.37 |
| 1320 | 0.85 | 0.65 | 0.75 | 0.75 | 0.08 | 1320 | 0.5 | 1.3 | 0.7 | 0.83 | 0.34 |
| 1380 | 0.85 | 0.65 | 0.8 | 0.77 | 0.08 | 1380 | 0.5 | 1.3 | 0.7 | 0.83 | 0.34 |
| 1440 | 0.85 | 0.65 | 0.8 | 0.77 | 0.08 | 1440 | 0.5 | 1.3 | 0.7 | 0.83 | 0.34 |
| 1500 | 0.85 | 0.65 | 0.8 | 0.77 | 0.08 | 1500 | 0.5 | 1.3 | 0.7 | 0.83 | 0.34 |
| 1560 | 0.85 | 0.65 | 0.8 | 0.77 | 0.08 | 1560 | 0.5 | 1.3 | 0.7 | 0.83 | 0.34 |
| 1620 | 0.85 | 0.65 | 0.8 | 0.77 | 0.08 | 1620 | 0.5 | 1.3 | 0.7 | 0.83 | 0.34 |
| 1680 | 0.85 | 0.65 | 0.8 | 0.77 | 0.08 | 1680 | 0.55 | 1.3 | 0.7 | 0.85 | 0.32 |
| 1740 | 0.85 | 0.65 | 0.8 | 0.77 | 0.08 | 1740 | 0.55 | 1.3 | 0.7 | 0.85 | 0.32 |
| 1800 | 0.85 | 0.65 | 0.8 | 0.77 | 0.08 | 1800 | 0.55 | 1.3 | 0.7 | 0.85 | 0.32 |

Table S2. Results of friction torque measurements in 60s intervals, their average values, and standard deviations for friction pairs with the POM-C component (continued).

| Polymer: Ertacetal C (POM-C); Lubricant: $C_{48}H_{102}O_4P_2$+CNT0.5% | | | | | | Polymer: Ertacetal C (POM-C); Lubricant: $C_{48}H_{102}O_4P_2$+CNT0.1% | | | | | |
|---|---|---|---|---|---|---|---|---|---|---|---|
| Time [s] | Friction torque [Nm] | | | | | Time [s] | Friction torque [Nm] | | | | |
| | Test 1 | Test 2 | Test 3 | Average | St. Dev. | | Test 1 | Test 2 | Test 3 | Average | St. Dev. |
| 60 | 0.3 | 0.35 | 0.35 | 0.33 | 0.02 | 60 | 0.35 | 0.3 | 0.35 | 0.33 | 0.02 |
| 120 | 0.65 | 0.6 | 0.6 | 0.62 | 0.02 | 120 | 0.7 | 0.6 | 0.7 | 0.67 | 0.05 |
| 180 | 0.95 | 0.9 | 0.9 | 0.92 | 0.02 | 180 | 1 | 0.95 | 1 | 0.98 | 0.02 |
| 240 | 0.95 | 0.9 | 0.95 | 0.93 | 0.02 | 240 | 0.9 | 0.9 | 1 | 0.93 | 0.05 |
| 300 | 0.9 | 0.8 | 0.9 | 0.87 | 0.05 | 300 | 0.9 | 0.9 | 0.9 | 0.90 | 0.00 |
| 360 | 0.9 | 0.8 | 0.9 | 0.87 | 0.05 | 360 | 0.9 | 0.9 | 0.9 | 0.90 | 0.00 |
| 420 | 0.85 | 0.8 | 0.9 | 0.85 | 0.04 | 420 | 0.85 | 0.85 | 0.85 | 0.85 | 0.00 |
| 480 | 0.85 | 0.8 | 0.9 | 0.85 | 0.04 | 480 | 0.85 | 0.85 | 0.8 | 0.83 | 0.02 |
| 540 | 0.85 | 0.8 | 0.9 | 0.85 | 0.04 | 540 | 0.85 | 0.85 | 0.8 | 0.83 | 0.02 |
| 600 | 0.8 | 0.8 | 0.9 | 0.83 | 0.05 | 600 | 0.85 | 0.9 | 0.8 | 0.85 | 0.04 |
| 660 | 0.8 | 0.8 | 0.9 | 0.83 | 0.05 | 660 | 0.85 | 0.9 | 0.8 | 0.85 | 0.04 |
| 720 | 0.8 | 0.8 | 0.9 | 0.83 | 0.05 | 720 | 0.85 | 0.9 | 0.85 | 0.87 | 0.02 |
| 780 | 0.8 | 0.8 | 0.9 | 0.83 | 0.05 | 780 | 0.85 | 0.9 | 0.85 | 0.87 | 0.02 |
| 840 | 0.8 | 0.8 | 0.9 | 0.83 | 0.05 | 840 | 0.85 | 0.9 | 0.85 | 0.87 | 0.02 |
| 900 | 0.75 | 0.8 | 0.85 | 0.80 | 0.04 | 900 | 0.85 | 0.9 | 0.85 | 0.87 | 0.02 |
| 960 | 0.75 | 0.8 | 0.85 | 0.80 | 0.04 | 960 | 0.85 | 0.9 | 0.85 | 0.87 | 0.02 |
| 1020 | 0.75 | 0.8 | 0.85 | 0.80 | 0.04 | 1020 | 0.85 | 0.9 | 0.85 | 0.87 | 0.02 |
| 1080 | 0.75 | 0.8 | 0.85 | 0.80 | 0.04 | 1080 | 0.85 | 0.9 | 0.85 | 0.87 | 0.02 |
| 1140 | 0.75 | 0.8 | 0.85 | 0.80 | 0.04 | 1140 | 0.9 | 0.9 | 0.85 | 0.88 | 0.02 |
| 1200 | 0.75 | 0.8 | 0.85 | 0.80 | 0.04 | 1200 | 0.9 | 0.85 | 0.85 | 0.87 | 0.02 |
| 1260 | 0.75 | 0.8 | 0.85 | 0.80 | 0.04 | 1260 | 0.9 | 0.85 | 0.85 | 0.87 | 0.02 |
| 1320 | 0.75 | 0.8 | 0.8 | 0.78 | 0.02 | 1320 | 0.9 | 0.85 | 0.85 | 0.87 | 0.02 |
| 1380 | 0.75 | 0.8 | 0.8 | 0.78 | 0.02 | 1380 | 0.9 | 0.85 | 0.85 | 0.87 | 0.02 |
| 1440 | 0.75 | 0.8 | 0.8 | 0.78 | 0.02 | 1440 | 0.9 | 0.85 | 0.85 | 0.87 | 0.02 |
| 1500 | 0.8 | 0.8 | 0.8 | 0.80 | 0.00 | 1500 | 0.9 | 0.85 | 0.85 | 0.87 | 0.02 |
| 1560 | 0.8 | 0.8 | 0.8 | 0.80 | 0.00 | 1560 | 0.9 | 0.85 | 0.85 | 0.87 | 0.02 |
| 1620 | 0.8 | 0.8 | 0.8 | 0.80 | 0.00 | 1620 | 0.9 | 0.85 | 0.85 | 0.87 | 0.02 |
| 1680 | 0.8 | 0.8 | 0.8 | 0.80 | 0.00 | 1680 | 0.9 | 0.85 | 0.85 | 0.87 | 0.02 |
| 1740 | 0.8 | 0.8 | 0.8 | 0.80 | 0.00 | 1740 | 0.9 | 0.85 | 0.85 | 0.87 | 0.02 |
| 1800 | 0.8 | 0.8 | 0.8 | 0.80 | 0.00 | 1800 | 0.9 | 0.85 | 0.85 | 0.87 | 0.02 |
| Polymer: Ertacetal C (POM-C); Lubricant: $C_{48}H_{102}O_4P_2$+Cu_CNT0.1% | | | | | | Polymer: Ertacetal C (POM-C); Lubricant: Reference | | | | | |
| Time [s] | Friction torque [Nm] | | | | | Time [s] | Friction torque [Nm] | | | | |
| | Test 1 | Test 2 | Test 3 | Average | St. Dev. | | Test 1 | Test 2 | Test 3 | Average | St. Dev. |
| 60 | 0.2 | 0.15 | 0.2 | 0.18 | 0.02 | 60 | 0.2 | 0.2 | 0.15 | 0.18 | 0.02 |
| 120 | 0.3 | 0.35 | 0.3 | 0.32 | 0.02 | 120 | 0.3 | 0.4 | 0.3 | 0.33 | 0.05 |
| 180 | 0.65 | 0.6 | 0.75 | 0.67 | 0.06 | 180 | 0.5 | 0.9 | 0.6 | 0.67 | 0.17 |
| 240 | 0.65 | 0.5 | 0.75 | 0.63 | 0.10 | 240 | 0.45 | 1.1 | 0.6 | 0.72 | 0.28 |
| 300 | 0.55 | 0.5 | 0.7 | 0.58 | 0.08 | 300 | 0.45 | 1.2 | 0.6 | 0.75 | 0.32 |
| 360 | 0.55 | 0.5 | 0.7 | 0.58 | 0.08 | 360 | 0.45 | 1.2 | 0.6 | 0.75 | 0.32 |
| 420 | 0.55 | 0.5 | 0.7 | 0.58 | 0.08 | 420 | 0.45 | 1.2 | 0.6 | 0.75 | 0.32 |
| 480 | 0.5 | 0.5 | 0.7 | 0.57 | 0.09 | 480 | 0.45 | 1.2 | 0.6 | 0.75 | 0.32 |
| 540 | 0.5 | 0.55 | 0.7 | 0.58 | 0.08 | 540 | 0.4 | 1.2 | 0.6 | 0.73 | 0.34 |
| 600 | 0.5 | 0.55 | 0.7 | 0.58 | 0.08 | 600 | 0.4 | 1.2 | 0.6 | 0.73 | 0.34 |
| 660 | 0.5 | 0.55 | 0.7 | 0.58 | 0.08 | 660 | 0.4 | 1.2 | 0.5 | 0.70 | 0.36 |
| 720 | 0.5 | 0.55 | 0.7 | 0.58 | 0.08 | 720 | 0.4 | 1.2 | 0.5 | 0.70 | 0.36 |
| 780 | 0.5 | 0.55 | 0.6 | 0.55 | 0.04 | 780 | 0.4 | 1.2 | 0.5 | 0.70 | 0.36 |
| 840 | 0.5 | 0.55 | 0.6 | 0.55 | 0.04 | 840 | 0.4 | 1.2 | 0.5 | 0.70 | 0.36 |
| 900 | 0.5 | 0.55 | 0.6 | 0.55 | 0.04 | 900 | 0.4 | 1.2 | 0.6 | 0.73 | 0.34 |
| 960 | 0.5 | 0.55 | 0.6 | 0.55 | 0.04 | 960 | 0.4 | 1.2 | 0.6 | 0.73 | 0.34 |
| 1020 | 0.5 | 0.55 | 0.6 | 0.55 | 0.04 | 1020 | 0.4 | 1.2 | 0.6 | 0.73 | 0.34 |
| 1080 | 0.5 | 0.55 | 0.6 | 0.55 | 0.04 | 1080 | 0.4 | 1.2 | 0.6 | 0.73 | 0.34 |
| 1140 | 0.5 | 0.55 | 0.6 | 0.55 | 0.04 | 1140 | 0.4 | 1.2 | 0.6 | 0.73 | 0.34 |
| 1200 | 0.5 | 0.55 | 0.6 | 0.55 | 0.04 | 1200 | 0.4 | 1.2 | 0.6 | 0.73 | 0.34 |
| 1260 | 0.5 | 0.55 | 0.6 | 0.55 | 0.04 | 1260 | 0.4 | 1.2 | 0.6 | 0.73 | 0.34 |
| 1320 | 0.5 | 0.55 | 0.65 | 0.57 | 0.06 | 1320 | 0.4 | 1.2 | 0.5 | 0.70 | 0.36 |
| 1380 | 0.5 | 0.55 | 0.65 | 0.57 | 0.06 | 1380 | 0.4 | 1.15 | 0.5 | 0.68 | 0.33 |
| 1440 | 0.5 | 0.55 | 0.65 | 0.57 | 0.06 | 1440 | 0.4 | 1.15 | 0.5 | 0.68 | 0.33 |
| 1500 | 0.5 | 0.55 | 0.65 | 0.57 | 0.06 | 1500 | 0.4 | 1.15 | 0.5 | 0.68 | 0.33 |
| 1560 | 0.5 | 0.55 | 0.65 | 0.57 | 0.06 | 1560 | 0.4 | 1.15 | 0.5 | 0.68 | 0.33 |
| 1620 | 0.5 | 0.55 | 0.65 | 0.57 | 0.06 | 1620 | 0.4 | 1.15 | 0.5 | 0.68 | 0.33 |
| 1680 | 0.5 | 0.55 | 0.65 | 0.57 | 0.06 | 1680 | 0.4 | 1.15 | 0.6 | 0.72 | 0.32 |
| 1740 | 0.5 | 0.55 | 0.65 | 0.57 | 0.06 | 1740 | 0.4 | 1.15 | 0.6 | 0.72 | 0.32 |
| 1800 | 0.5 | 0.55 | 0.65 | 0.57 | 0.06 | 1800 | 0.4 | 1.15 | 0.6 | 0.72 | 0.32 |

Table S3. Results of friction torque measurements in 60s intervals, their average values, and standard deviations for friction pairs with the PA component.

| Polymer: Ertalon 6SA (PA); Lubricant: $C_{48}H_{102}O_4P_2$ | | | | | | Polymer: Ertalon 6SA (PA); Lubricant: $C_{48}H_{102}O_4P_2$+CNT1% | | | | | |
|---|---|---|---|---|---|---|---|---|---|---|---|
| Time [s] | Friction torque [Nm] | | | | | Time [s] | Friction torque [Nm] | | | | |
| | Test 1 | Test 2 | Test 3 | Average | St. Dev. | | Test 1 | Test 2 | Test 3 | Average | St. Dev. |
| 60 | 0.15 | 0.2 | 0.2 | 0.18 | 0.02 | 60 | 0.2 | 0.2 | 0.15 | 0.18 | 0.02 |
| 120 | 0.45 | 0.5 | 0.45 | 0.47 | 0.02 | 120 | 0.4 | 0.5 | 0.4 | 0.43 | 0.05 |
| 180 | 0.7 | 0.85 | 0.8 | 0.78 | 0.06 | 180 | 0.75 | 0.85 | 0.65 | 0.75 | 0.08 |
| 240 | 0.7 | 0.85 | 0.8 | 0.78 | 0.06 | 240 | 0.8 | 0.85 | 0.65 | 0.77 | 0.08 |
| 300 | 0.7 | 0.9 | 0.8 | 0.80 | 0.08 | 300 | 0.8 | 0.85 | 0.65 | 0.77 | 0.08 |
| 360 | 0.7 | 0.9 | 0.8 | 0.80 | 0.08 | 360 | 0.75 | 0.85 | 0.65 | 0.75 | 0.08 |
| 420 | 0.7 | 0.9 | 0.8 | 0.80 | 0.08 | 420 | 0.75 | 0.85 | 0.65 | 0.75 | 0.08 |
| 480 | 0.75 | 0.9 | 0.9 | 0.85 | 0.07 | 480 | 0.75 | 0.85 | 0.6 | 0.73 | 0.10 |
| 540 | 0.75 | 0.9 | 0.9 | 0.85 | 0.07 | 540 | 0.75 | 0.8 | 0.6 | 0.72 | 0.08 |
| 600 | 0.75 | 0.9 | 0.9 | 0.85 | 0.07 | 600 | 0.75 | 0.8 | 0.6 | 0.72 | 0.08 |
| 660 | 0.75 | 0.9 | 0.95 | 0.87 | 0.08 | 660 | 0.75 | 0.8 | 0.6 | 0.72 | 0.08 |
| 720 | 0.7 | 0.9 | 0.95 | 0.85 | 0.11 | 720 | 0.7 | 0.75 | 0.55 | 0.67 | 0.08 |
| 780 | 0.7 | 0.9 | 0.95 | 0.85 | 0.11 | 780 | 0.7 | 0.75 | 0.55 | 0.67 | 0.08 |
| 840 | 0.7 | 0.9 | 0.95 | 0.85 | 0.11 | 840 | 0.7 | 0.7 | 0.55 | 0.65 | 0.07 |
| 900 | 0.65 | 0.9 | 0.95 | 0.83 | 0.13 | 900 | 0.7 | 0.7 | 0.5 | 0.63 | 0.09 |
| 960 | 0.65 | 0.9 | 0.95 | 0.83 | 0.13 | 960 | 0.7 | 0.7 | 0.5 | 0.63 | 0.09 |
| 1020 | 0.65 | 0.9 | 1 | 0.85 | 0.15 | 1020 | 0.7 | 0.7 | 0.5 | 0.63 | 0.09 |
| 1080 | 0.65 | 0.95 | 1 | 0.87 | 0.15 | 1080 | 0.7 | 0.65 | 0.45 | 0.60 | 0.11 |
| 1140 | 0.65 | 0.95 | 1 | 0.87 | 0.15 | 1140 | 0.65 | 0.65 | 0.45 | 0.58 | 0.09 |
| 1200 | 0.65 | 0.95 | 1 | 0.87 | 0.15 | 1200 | 0.65 | 0.65 | 0.45 | 0.58 | 0.09 |
| 1260 | 0.65 | 0.95 | 1 | 0.87 | 0.15 | 1260 | 0.65 | 0.65 | 0.45 | 0.58 | 0.09 |
| 1320 | 0.65 | 0.95 | 1 | 0.87 | 0.15 | 1320 | 0.65 | 0.65 | 0.45 | 0.58 | 0.09 |
| 1380 | 0.65 | 0.95 | 1 | 0.87 | 0.15 | 1380 | 0.65 | 0.65 | 0.45 | 0.58 | 0.09 |
| 1440 | 0.65 | 0.95 | 1 | 0.87 | 0.15 | 1440 | 0.65 | 0.65 | 0.45 | 0.58 | 0.09 |
| 1500 | 0.65 | 0.9 | 1.05 | 0.87 | 0.16 | 1500 | 0.65 | 0.65 | 0.45 | 0.58 | 0.09 |
| 1560 | 0.65 | 0.9 | 1.05 | 0.87 | 0.16 | 1560 | 0.65 | 0.65 | 0.45 | 0.58 | 0.09 |
| 1620 | 0.65 | 0.9 | 1.05 | 0.87 | 0.16 | 1620 | 0.65 | 0.65 | 0.45 | 0.58 | 0.09 |
| 1680 | 0.65 | 0.95 | 1.05 | 0.88 | 0.17 | 1680 | 0.65 | 0.65 | 0.45 | 0.58 | 0.09 |
| 1740 | 0.65 | 0.95 | 1.05 | 0.88 | 0.17 | 1740 | 0.65 | 0.65 | 0.45 | 0.58 | 0.09 |
| 1800 | 0.65 | 0.95 | 1.05 | 0.88 | 0.17 | 1800 | 0.65 | 0.65 | 0.45 | 0.58 | 0.09 |
| Polymer: Ertalon 6SA (PA); Lubricant: $C_{48}H_{102}O_4P_2$+CNT0.5% | | | | | | Polymer: Ertalon 6SA (PA); Lubricant: $C_{48}H_{102}O_4P_2$+CNT0.1% | | | | | |
| Time [s] | Friction torque [Nm] | | | | | Time [s] | Friction torque [Nm] | | | | |
| | Test 1 | Test 2 | Test 3 | Average | St. Dev. | | Test 1 | Test 2 | Test 3 | Average | St. Dev. |
| 60 | 0.2 | 0.2 | 0.25 | 0.22 | 0.02 | 60 | 0.2 | 0.2 | 0.3 | 0.23 | 0.05 |
| 120 | 0.55 | 0.45 | 0.45 | 0.48 | 0.05 | 120 | 0.45 | 0.5 | 0.55 | 0.50 | 0.04 |
| 180 | 0.95 | 0.7 | 0.9 | 0.85 | 0.11 | 180 | 0.75 | 0.8 | 0.85 | 0.80 | 0.04 |
| 240 | 0.95 | 0.7 | 0.9 | 0.85 | 0.11 | 240 | 0.75 | 0.8 | 0.85 | 0.80 | 0.04 |
| 300 | 0.95 | 0.7 | 0.9 | 0.85 | 0.11 | 300 | 0.75 | 0.8 | 0.9 | 0.82 | 0.06 |
| 360 | 0.95 | 0.65 | 0.9 | 0.83 | 0.13 | 360 | 0.75 | 0.8 | 0.9 | 0.82 | 0.06 |
| 420 | 1 | 0.65 | 0.9 | 0.85 | 0.15 | 420 | 0.75 | 0.8 | 1 | 0.85 | 0.11 |
| 480 | 1 | 0.65 | 0.85 | 0.83 | 0.14 | 480 | 0.75 | 0.8 | 1 | 0.85 | 0.11 |
| 540 | 1 | 0.65 | 0.85 | 0.83 | 0.14 | 540 | 0.75 | 0.8 | 1 | 0.85 | 0.11 |
| 600 | 1.05 | 0.7 | 0.85 | 0.87 | 0.14 | 600 | 0.75 | 0.8 | 1 | 0.85 | 0.11 |
| 660 | 1.05 | 0.7 | 0.85 | 0.87 | 0.14 | 660 | 0.75 | 0.8 | 1 | 0.85 | 0.11 |
| 720 | 1.05 | 0.7 | 0.85 | 0.87 | 0.14 | 720 | 0.75 | 0.8 | 1 | 0.85 | 0.11 |
| 780 | 1.1 | 0.65 | 0.85 | 0.87 | 0.18 | 780 | 0.8 | 0.8 | 1.05 | 0.88 | 0.12 |
| 840 | 1.1 | 0.65 | 0.85 | 0.87 | 0.18 | 840 | 0.8 | 0.8 | 1.05 | 0.88 | 0.12 |
| 900 | 1.1 | 0.65 | 0.85 | 0.87 | 0.18 | 900 | 0.8 | 0.9 | 1.05 | 0.92 | 0.10 |
| 960 | 1.1 | 0.65 | 0.85 | 0.87 | 0.18 | 960 | 0.8 | 0.9 | 1.1 | 0.93 | 0.12 |
| 1020 | 1.1 | 0.65 | 0.85 | 0.87 | 0.18 | 1020 | 0.8 | 0.9 | 1.1 | 0.93 | 0.12 |
| 1080 | 1.1 | 0.6 | 0.85 | 0.85 | 0.20 | 1080 | 0.8 | 0.9 | 1.1 | 0.93 | 0.12 |
| 1140 | 1.1 | 0.6 | 0.8 | 0.83 | 0.21 | 1140 | 0.85 | 0.95 | 1.1 | 0.97 | 0.10 |
| 1200 | 1.1 | 0.6 | 0.8 | 0.83 | 0.21 | 1200 | 0.85 | 0.95 | 1.1 | 0.97 | 0.10 |
| 1260 | 1.1 | 0.6 | 0.8 | 0.83 | 0.21 | 1260 | 0.85 | 0.95 | 1.1 | 0.97 | 0.10 |
| 1320 | 1.05 | 0.6 | 0.8 | 0.82 | 0.18 | 1320 | 0.85 | 0.95 | 1.1 | 0.97 | 0.10 |
| 1380 | 1.05 | 0.6 | 0.8 | 0.82 | 0.18 | 1380 | 0.85 | 0.95 | 1.1 | 0.97 | 0.10 |
| 1440 | 1.05 | 0.6 | 0.8 | 0.82 | 0.18 | 1440 | 0.85 | 0.95 | 1.1 | 0.97 | 0.10 |
| 1500 | 1.05 | 0.6 | 0.8 | 0.82 | 0.18 | 1500 | 0.85 | 0.95 | 1.1 | 0.97 | 0.10 |
| 1560 | 1.05 | 0.6 | 0.8 | 0.82 | 0.18 | 1560 | 0.85 | 0.95 | 1.1 | 0.97 | 0.10 |
| 1620 | 1.05 | 0.6 | 0.8 | 0.82 | 0.18 | 1620 | 0.85 | 0.95 | 1.1 | 0.97 | 0.10 |
| 1680 | 1.05 | 0.6 | 0.8 | 0.82 | 0.18 | 1680 | 0.95 | 0.95 | 1.1 | 1.00 | 0.07 |
| 1740 | 1.05 | 0.6 | 0.8 | 0.82 | 0.18 | 1740 | 0.95 | 0.95 | 1.1 | 1.00 | 0.07 |
| 1800 | 1.05 | 0.6 | 0.8 | 0.82 | 0.18 | 1800 | 0.95 | 0.95 | 1.1 | 1.00 | 0.07 |

Table S3. Results of friction torque measurements in 60s intervals, their average values, and standard deviations for friction pairs with the PA component (continued).

| | Polymer: Ertalon 6SA (PA); Lubricant: $C_{48}H_{102}O_4P_2$+Cu_CNT0.1% | | | | | | Polymer: Ertalon 6SA (PA); Lubricant: Reference | | | | |
|---|---|---|---|---|---|---|---|---|---|---|---|
| Time [s] | Friction torque [Nm] | | | | | Time [s] | Friction torque [Nm] | | | | |
| | Test 1 | Test 2 | Test 3 | Average | St. Dev. | | Test 1 | Test 2 | Test 3 | Average | St. Dev. |
| 60 | 0.1 | 0.1 | 0.1 | 0.10 | 0.00 | 60 | 0.2 | 0.15 | 0.15 | 0.17 | 0.02 |
| 120 | 0.2 | 0.3 | 0.2 | 0.23 | 0.05 | 120 | 0.5 | 0.4 | 0.55 | 0.48 | 0.06 |
| 180 | 0.65 | 0.45 | 0.45 | 0.52 | 0.09 | 180 | 1.05 | 0.75 | 1 | 0.93 | 0.13 |
| 240 | 0.65 | 0.5 | 0.45 | 0.53 | 0.08 | 240 | 1.05 | 0.75 | 1.1 | 0.97 | 0.15 |
| 300 | 0.65 | 0.5 | 0.45 | 0.53 | 0.08 | 300 | 1.1 | 0.75 | 1.1 | 0.98 | 0.16 |
| 360 | 0.65 | 0.5 | 0.45 | 0.53 | 0.08 | 360 | 1.1 | 0.75 | 1.2 | 1.02 | 0.19 |
| 420 | 0.7 | 0.55 | 0.45 | 0.57 | 0.10 | 420 | 1.1 | 0.75 | 1.2 | 1.02 | 0.19 |
| 480 | 0.7 | 0.55 | 0.5 | 0.58 | 0.08 | 480 | 1.1 | 0.85 | 1.2 | 1.05 | 0.15 |
| 540 | 0.7 | 0.55 | 0.5 | 0.58 | 0.08 | 540 | 1.1 | 0.85 | 1.2 | 1.05 | 0.15 |
| 600 | 0.7 | 0.55 | 0.5 | 0.58 | 0.08 | 600 | 1.15 | 0.85 | 0.75 | 0.92 | 0.17 |
| 660 | 0.7 | 0.65 | 0.6 | 0.65 | 0.04 | 660 | 1.15 | 0.85 | 0.75 | 0.92 | 0.17 |
| 720 | 0.7 | 0.65 | 0.6 | 0.65 | 0.04 | 720 | 1.15 | 0.85 | 1.4 | 1.13 | 0.22 |
| 780 | 0.7 | 0.65 | 0.6 | 0.65 | 0.04 | 780 | 1.3 | 0.85 | 1.4 | 1.18 | 0.24 |
| 840 | 0.7 | 0.65 | 0.6 | 0.65 | 0.04 | 840 | 1.3 | 0.95 | 1.4 | 1.22 | 0.19 |
| 900 | 0.75 | 0.65 | 0.6 | 0.67 | 0.06 | 900 | 1.3 | 0.95 | 1.4 | 1.22 | 0.19 |
| 960 | 0.75 | 0.65 | 0.6 | 0.67 | 0.06 | 960 | 1.35 | 0.95 | 1.4 | 1.23 | 0.20 |
| 1020 | 0.75 | 0.7 | 0.6 | 0.68 | 0.06 | 1020 | 1.35 | 0.95 | 1.4 | 1.23 | 0.20 |
| 1080 | 0.75 | 0.7 | 0.6 | 0.68 | 0.06 | 1080 | 1.35 | 0.95 | 1.4 | 1.23 | 0.20 |
| 1140 | 0.75 | 0.7 | 0.6 | 0.68 | 0.06 | 1140 | 1.35 | 0.95 | 1.4 | 1.23 | 0.20 |
| 1200 | 0.75 | 0.7 | 0.6 | 0.68 | 0.06 | 1200 | 1.35 | 0.95 | 1.4 | 1.23 | 0.20 |
| 1260 | 0.75 | 0.7 | 0.6 | 0.68 | 0.06 | 1260 | 1.5 | 0.95 | 1.4 | 1.28 | 0.24 |
| 1320 | 0.75 | 0.7 | 0.6 | 0.68 | 0.06 | 1320 | 1.5 | 1.1 | 1.4 | 1.33 | 0.17 |
| 1380 | 0.75 | 0.7 | 0.6 | 0.68 | 0.06 | 1380 | 1.4 | 1.1 | 1.4 | 1.30 | 0.14 |
| 1440 | 0.75 | 0.7 | 0.75 | 0.73 | 0.02 | 1440 | 1.4 | 1.1 | 1.5 | 1.33 | 0.17 |
| 1500 | 0.75 | 0.7 | 0.75 | 0.73 | 0.02 | 1500 | 1.3 | 1.1 | 1.5 | 1.30 | 0.16 |
| 1560 | 0.75 | 0.7 | 0.75 | 0.73 | 0.02 | 1560 | 1.3 | 1.1 | 1.5 | 1.30 | 0.16 |
| 1620 | 0.75 | 0.7 | 0.75 | 0.73 | 0.02 | 1620 | 1.3 | 1.15 | 1.5 | 1.32 | 0.14 |
| 1680 | 0.75 | 0.7 | 0.75 | 0.73 | 0.02 | 1680 | 1.3 | 1.15 | 1.5 | 1.32 | 0.14 |
| 1740 | 0.75 | 0.7 | 0.75 | 0.73 | 0.02 | 1740 | 1.3 | 1.15 | 1.5 | 1.32 | 0.14 |
| 1800 | 0.75 | 0.7 | 0.75 | 0.73 | 0.02 | 1800 | 1.3 | 1.15 | 1.5 | 1.32 | 0.14 |